\documentclass{article}
\usepackage{amssymb}
\usepackage{amsmath}
\usepackage{amsfonts}

\newtheorem{theorem}{Theorem}

\newtheorem{notation}[theorem]{Notation}

\input{tcilatex}
\begin{document}

\title{\textbf{From Grad-Shafranov Equations set to a pseudo-general form of
the }\\
\textbf{Non-Linear Schr\"{o}dinger Equation}\\
\ \ \ \ \ \ \ \ \ \ \ \ \ \ \ \ \ \ \ \ \ \ \ \ \ \ \ \ \ \ \ \ \ \ \ \ \ \
\ \ \ \ \ \ \ \ \ \ \ \ \ \ \ \ \ \ \ \ \ \ \ \ \ \ \ \ \ \ \ \ \ \ \ \ \ \
\ \ \ \ \ \ \ \ \ \ \ \ \ \ \ \ \ \ \ \ \ \ \ \ \ \ \ \ \ \ \ \ \ \ \ \\
\textsl{About the validity conditions of some general transformation
functionals of the free fields in the Grad-Shafranov Equations}}
\author{\textit{M.Romeo }(\textit{\thanks{%
( permanent e-mail address: \textit{michele.romeo.mr@gmail.com })} }) \\
\\
\textit{Theoretical Sector, Department of Physics, University of Salento, }\\
\textit{Lecce (Italy) }}
\maketitle

\begin{abstract}
In the year 2003 a paper by \textit{G. Lapenta} demonstrated that there is a
\textquotedblleft new class of soliton-like solutions for the Grad-Shafranov
Equations (GSE)\textquotedblright . The author determined an appropriate
pair of transformations of the free fields $\mathit{p}$ (fluid field of
hydrodynamical pressure) and $\mathit{B}_{z}$ (z-component of magnetic
induction field) that leads from the Helmholtz Equation to the Non-Linear
Schr\"{o}dinger Equation (NLSE) with cubic non-linearity. In the following
year (2004), the work of Lapenta was opposed by \textit{G.N.
Throumoulopoulos et al.}, who criticized his idea\ \ of the field
transformations as a mathematically incoherent choice; contextually the
authors suggested a new point of view for this one. In his response, in the
same year, \textit{G. Lapenta} carried out numerical simulations that showed
the existence of solitonic structures in a Magnetohydrodynamical (MHD)
plasma-context. In the present work I want to demonstrate a \textit{critical
condition} for a 'complex' poloidal flux function $\mathit{\Psi }_{p}$ in a
plane framework $\mathit{(x,y)}$ that leads to a class of pseudo-general
NLSEs which establishes the validity of both \textit{G.Lapenta} and \textit{%
G.N. Throumoulopoulos et al.} choices for the field transformation
functionals.\ \ \ \ \ \ \ \ \ \ \ \ \ \ \ \ \ \ \ \ \ \ \ \ \ \ \ \ \ \ \ \
\ \ \ \ \ \ \ \ \ \ 
\end{abstract}

\section{\protect\Large Introduction}

{\large In theoretical models in which the accretion dynamics in
magnetohydrodynamical plasmas are explored ,i.e. plasmas that obey at the
various manetohydrodynamical models ,with good approximation, in Plasma
Physics, like resistive model, Hall-model and others mixed and dominant
models which include the General-Relativistic Theory, we can see that the
astrophysical plasmas develop themselves around compact objects, \ like
neutron stars and black holes usually; in such theoretical models in which
the formations of anti-parallel plasma-jets are often examined in the
context of their singular morphology, there are many references at well
defined characteristic mathematical structures that could explain the shape
of some sub-formation inside the accretion and ejection plasma mechanics.}

{\large These mathematical models result very interesting for both their
internal coherence and physical plausibility, because they can impose an
indiscutible mathematical link between the fundamental theory of the above
astrophysical phenomena and the physical charateristics of the singular
plasma formations in the above star structures, without affecting the
essential set of solutions in the base phisical models that explain very
well yet the most considerable star dynamics in the astrophysical plasma
context.}

{\large For example, there are many mathematical singular structures often
investigated because they match at several characteristic cluster formations
inside the accretion disc and the plasma jets; these cluster formations are
well localized physical structures, with magnetic induction field }$\mathbf{B%
}${\large \ and electric field }$\mathbf{E}$ {\large topologies that take
place in the surrounding space and that donate, with the necessary
hydrodynamical force fields, a precise shape at these formations, which are
therefore confined in limited space areas in the curved space around the
concerned compact object (this situation exist because the gravitational
field densities near these astrophysical objects are relevant and a full
relativistic treatment in any possible theoretical model must be considered)
and that show in many cases a peculiar space periodicity in a well defined
global space configuration when the magnetohydrodynamical mechanics are in
steady state condition, or in an equilibrium state.}

{\large These localized plasma formation are indentifiable, for example, in
the space development of a plasma jet which derives from a compact star and
it stretch itself from the central object to the far space areas by a series
of quasi-spherical plasma structures named 'plasma bubbles' \cite{1},\cite{2}
(in this case the jet acquired a singular morphology known by the name of \
'knotty structure') .}

{\large In other cases, similar structures are localized in the accretion
discs around same compact stars, leading to the formation of related
structures morphologically, in which the characteristic 'toroidal' sha- pe
is linked to the spiral motions in the disc, which response to a specifical
magnetohydrodynamical equations set of equlibrium state that persist on a
long time-scale, meaning in this way the same steady-state configuration. In
this regard, localized structures in the dynamics of accretion discs in an
astrophysical context have been identified analytically and
'experimentally', by several numerical simulations, from }\textit{%
Petviashvili et al. }\cite{3},\cite{4}\textit{\ }{\large in the 80s as a
solution to the problem of the Grad-Shafranov Equation (GSE), a fundamental
non-linear equilibrium equation in Plasma Physics, by the context of a
general problem of two-dimensional dynamical plasma equilibrium.}

{\large Indeed} {\large the Grad-Shafranov Equation \footnote{%
( H. Grad and H. Rubin (1958), V.D. Shafranov (1966) )} is a 2nd-order
elliptic PDE, from the mathematical point of view, which responds to a
plasma equlibrium problem in a two-dimensional framework }$(x,y)${\large \ 
\cite{14}\ and it's derived by the Euler Equation (EE) in the ideal limit
(ideal Magnetohydrodynamics) and the Ampere's Law (i.e. 'source equation' in
the Maxwell's set equations) with neglect the 'displacement current',
because the non-relativistic limit }$(v<<c)$ {\large or the quasi-classic
relativistic limit (it takes into account by a 'pseudo-newtonian'
gravitational potential, like Paczynski-Wiita potential) of the gravity
potential term in the Euler Equation, which derived for the concerned
hydrodynamical problem (i.e. accretion disc or plasma jet in the
astrophysical objects framework).}

{\large \ This equation establish therefore a 'non-linear problem'}\textit{\ 
}{\large for a certain scalar 'flux function' }$\mathit{\Psi }$ {\large %
which defines the induction field }$\boldsymbol{B}$ {\large in the ordinary
space }$(x,y,z)$ {\large around the compact star and which have been yet
analized from }\textit{Petviashvili et al. }\cite{3}\textit{\ }{\large in an
analytical work that demonstrated the existence of \textit{solitary toroidal
structures} in the validity framework of the GSE;}

{\large furthermore, the GSE for this flux function }$\mathit{\Psi }$ 
{\large establish the necessary conditions (i.e. a 'critic conditions') to
which the 'parameters space' and the free fields in the fluid equations must
obey so that the equilibrium scenario is respected in the steady-state
dynamics.}\bigskip

{\large At this point, it's indispensable to watch that in many non-linear
problems of Mathematical-Physics that usually develop some sets of
expressions containing PDEs, like non linear wave equations, we are led to
consider, by an appropriate choice of a function transformations set and a
'reduction of potentials', an equivalent and more general non-linear
differential problem that represents the 'root differential problem' for a
more general category of non-linear problems. From the point of view of this
one, the more considerable non-linear equation that is present in many
problems of Mathematical-Physics regarding non-linear wave phenomena that
involve PDEs is obviously the well known '}\textit{Non-Linear Schr\"{o}%
dinger Equation' \ }{\large (NLSE), an equation that is often present, for
example, in non-linear optical problems \cite{11}.}

{\large You know the best known solution scenario of this non-linear
equation is a '\textit{solitonic solution}' or '\textit{soliton solution}',
where a 'soliton'}\textit{\ }{\large is the well known special solitary wave
solution of several non-linear dispersive wave equations; the main
characteristics of this wave solution are its singular mathematical
structure and its physical behavior, which synthesize them primarily in
their peculiarity of \textit{localized wave-structure}. }

{\large It seems therefore that the GSE, a non-linear (elliptic) equation,
can be lead to the more general form of the NLSE for the stationary case
(steady-state for a plasma in an equlibrium state) by an appropriate choice
of 'transformation functionals'}\textit{\ }{\large for the free fields
inside it; this consideration is furthermore reasoned by the fact that the
peculiar plasma \textit{knots} in the accretion discs and plasma jets cited
above could be justified by an appropriate 're-interpretation' of these
formations as periodic solitonic structures with justified contour
conditions.}

{\large At this point, if such appropriate choice of a transformations set
exist and it is coherent from the mathematical point of view, there are no
reasons for to refuse such a possibility. Such a choice for turning GSE in a
NLSE with a specific cubic non-linearity, i.e. the NLSE for the \textit{Kerr
effect} in a non-linear optical mean, was made recently by }\textit{%
G.Lapenta }\cite{8}{\large , who was demonstrated how it's possible to
obtain a pair of appropriate transformations in a two-dimensional framework }%
$(x,y)$, {\large like that in the equatorial plane of an accretion disc,} 
{\large which lead from a general GSE to a cubic NLSE by a two-dimensional
Helmholtz Equation (2dHE) \cite{7}, which is important in the non-linear
optical problems and it is related to many problems for steady-state
oscillations (mechanical, acoustical, thermal, electromagnetic); but in the
following year, by a purely mathematical work, }\textit{Lapenta}{\large 
\textit{'}s idea was opposed by }\textit{Throumoulopoulos G.N., Hizanidis K.
and Tasso H. }\cite{9}\textit{\large , }{\large whom criticized one of his
pair of free fields transformations because its supposed mathematical
inconsistency. The authors proposed furthermore a new pair of correct
transformations for the same problem.}

{\large In this work both }\textit{Lapenta G.}{\large \ and }\textit{%
Throumoulopoulos G.N. et al.}{\large \ positions will be examined and a new
and more general choice will be proposed by me, who will show that a
'critical condition' exist for the flux function }$\mathit{\Psi }$, {\large %
so that this general choice be a coherent choice}, {\large and} {\large both
choices are not only mathematically coherent but the more general choice
cited above include any possible set of reasonable transformations, leading
in this way to a more desiderable pseudo-general form of the NLSE \cite{6}.}

{\large All the analytical considerations will be made for the case of
knotty structures in the accretion discs as in }\textit{Lapenta}{\large 's
work, therefore in the equatorial plane or in the }$(x,y)$ {\large %
framework; in this case }$z$ {\large is a negligible variable.}

{\large A brief} {\large mention to the 'moment method' in general NLSE
examined in the paper of }\textit{Garc\'{\i}a-Ripoll, J.J. and P\'{e}rez-Garc%
\'{\i}a, V.M. }\cite{6}\textit{\ }{\large will be done in the next paragraph.%
}

\section{\protect\Large From GSEs to a pseudo-general form of the NLSE -
stationary case}

{\large We derive in the next sections a general choice for the
transformation functionals of the free fields in the GSEs which leads to a
named pseudo-general form of the NLSE in the stationary case (steady-state
case), showing that this choice include both }\textit{Lapenta G.}{\large \
and }\textit{Throumoulopoulos G.N. et al.}{\large \ choices and putting in
evidence that the same choices are compatible with one another.}

\subsection{\protect\Large Grad-Shafranov Equations set\newline
}

{\large The GSEs set, in the more general form for a fully ionized plasma }$%
(n_{e}=Zn_{i})${\large , can be derive by both the Euler Equations in the
ideal limit for the ions and the electrons in a two-fluid treatment and the
Ampere's Law; in this respect, we take the two-fluid equations cited in }%
\textit{Lighthill, M.J. }\cite{13}$\ \ \ \ \ \ \ \ \ $%
\begin{equation}
{\Large n}_{e}{\Large m}_{e}\left( \frac{{\LARGE \partial }}{{\LARGE %
\partial t}}\mathbf{v}_{e}+\mathbf{v}_{e}\cdot {\Large \nabla }\mathbf{v}%
_{e}\right) =-{\Large \nabla }{\large p}_{e}-\mathbf{M}-{\Large n}_{e}%
{\Large e(}\mathbf{E+v}_{e}\times \mathbf{B}{\Large )}  \tag{2.1}
\label{2.1}
\end{equation}

$\ \ \ \ \ \ \ $%
\begin{equation}
\ {\Large n}_{i}{\Large m}_{i}\left( \frac{{\LARGE \partial }}{{\LARGE %
\partial t}}\mathbf{v}_{i}+\mathbf{v}_{i}\cdot {\Large \nabla }\mathbf{v}%
_{i}\right) =-{\Large \nabla }{\large p}_{i}+\mathbf{M}+{\Large n}_{i}%
{\Large Ze(}\mathbf{E+v}_{i}\times \mathbf{B}{\Large )}  \tag{2.2}
\label{2.2}
\end{equation}%
{\large where }$\mathbf{M}$ {\large is the rate of loss of electron
momentum, per unit volume, by collisions with ions; neglecting electrons and
ions inertia (\textit{non-inertial} approximation) and adding both
corresponding sides of (2.1) and (2.2), we obtain}%
\begin{eqnarray*}
\mathbf{0} &=&-{\Large \nabla }{\large p}_{e}-{\Large \nabla }{\large p}_{i}-%
{\Large n}_{e}{\Large e(}\mathbf{E}+\mathbf{v}_{e}\times \mathbf{B}{\Large )}%
+{\Large n}_{i}{\Large Ze(}\mathbf{E}+\mathbf{v}_{i}\times \mathbf{B}{\Large %
)}= \\
&=&-{\Large \nabla }{\large p}+{\large n}_{e}{\large e(}\mathbf{v}_{i}-%
\mathbf{v}_{e}{\large )}\times \mathbf{B=}-{\Large \nabla }{\large p}+%
\mathbf{J}\times \mathbf{B}
\end{eqnarray*}%
{\large which lead to}%
\begin{equation}
{\Large \nabla }{\large p\ }=\mathbf{J}\times \mathbf{B}  \tag{2.3}
\label{2.3}
\end{equation}%
{\large being the current density}%
\begin{equation*}
\mathbf{J}={\large n}_{e}{\large e(}\mathbf{v}_{i}-\mathbf{v}_{e}{\large )}
\end{equation*}%
{\large At this point the calculus is the same of }\textit{Lapenta}{\large %
's work. Assuming a plane framework }$(x,y)$ {\large because} {\large the
analysis }\bigskip {\large is made on the equatorial plane of an accretion
disc,we can take as a particular solution of (2.3) \cite{12}}%
\begin{equation}
\mathbf{B}=\widehat{\mathbf{z}}\times {\Large \nabla }\Psi +B_{z}\widehat{%
\mathbf{z}}  \tag{2.4}  \label{2.4}
\end{equation}%
{\large The Ampere's Law lead by the (2.4) to}%
\begin{eqnarray*}
\mathbf{J} &=&{\Large \nabla }\times \mathbf{B}=\widehat{\mathbf{z}}\left( 
{\Large \nabla }\cdot {\Large \nabla }\Psi \right) -{\Large \nabla }\Psi
\left( {\Large \nabla }\cdot \widehat{\mathbf{z}}\right) +\left( {\Large %
\nabla }\Psi \cdot {\Large \nabla }\right) \widehat{\mathbf{z}}+ \\
&&-\left( \widehat{\mathbf{z}}\cdot {\Large \nabla }\right) {\Large \nabla }%
\Psi +B_{z}\left( {\Large \nabla }\times \widehat{\mathbf{z}}\right) -%
\widehat{\mathbf{z}}\times {\Large \nabla }B_{z}
\end{eqnarray*}%
{\large which delivers, being }$z$ {\large a negligible variable,}%
\begin{equation}
\mathbf{J}=\widehat{\mathbf{z}}{\Large \nabla }^{2}\Psi -\widehat{\mathbf{z}}%
\times {\Large \nabla }B_{z}  \tag{2.5}  \label{2.5}
\end{equation}%
{\large If we use the current density expression (2.5) and the solution
(2.4) in (2.3), we obtain the equation}%
\begin{eqnarray*}
{\Large \nabla }{\large p\ } &=&\left( \widehat{\mathbf{z}}{\Large \nabla }%
^{2}\Psi -\widehat{\mathbf{z}}\times {\Large \nabla }B_{z}\right) \times
\left( \widehat{\mathbf{z}}\times {\Large \nabla }\Psi +B_{z}\widehat{%
\mathbf{z}}\right) = \\
&=&\widehat{\mathbf{z}}{\Large \nabla }^{2}\Psi \times \left( \widehat{%
\mathbf{z}}\times {\Large \nabla }\Psi \right) -\left( \widehat{\mathbf{z}}%
\times {\Large \nabla }B_{z}\right) \times \left( \widehat{\mathbf{z}}\times 
{\Large \nabla }\Psi \right) -\left( \widehat{\mathbf{z}}\times {\Large %
\nabla }B_{z}\right) \times B_{z}\widehat{\mathbf{z}}= \\
&=&{\Large \nabla }^{2}\Psi \left( \widehat{\mathbf{z}}\widehat{\mathbf{z}}%
\cdot {\Large \nabla }\Psi -{\Large \nabla }\Psi \widehat{\mathbf{z}}\cdot 
\widehat{\mathbf{z}}\right) -\left[ \left( \widehat{\mathbf{z}}\times 
{\Large \nabla }B_{z}\cdot {\Large \nabla }\Psi \right) \widehat{\mathbf{z}}%
-\left( \widehat{\mathbf{z}}\times {\Large \nabla }B_{z}\cdot \widehat{%
\mathbf{z}}\right) {\Large \nabla }\Psi \right] + \\
&&+B_{z}\left( \widehat{\mathbf{z}}\widehat{\mathbf{z}}\cdot {\Large \nabla }%
B_{z}-{\Large \nabla }B_{z}\widehat{\mathbf{z}}\cdot \widehat{\mathbf{z}}%
\right)
\end{eqnarray*}%
{\large which delivers}%
\begin{equation}
{\Large \nabla }{\large p\ }+{\Large \nabla }^{2}\Psi {\Large \nabla }\Psi
+\left( \widehat{\mathbf{z}}\times {\Large \nabla }B_{z}\cdot {\Large \nabla 
}\Psi \right) \widehat{\mathbf{z}}+B_{z}{\Large \nabla }B_{z}=\mathbf{0} 
\tag{2.6}  \label{2.6}
\end{equation}%
{\large Observing that equation (2.6) is made of two parts linearly
independent among them, from this one it follows the equations system}%
\begin{equation*}
\left( \widehat{\mathbf{z}}\times {\Large \nabla }B_{z}\cdot {\Large \nabla }%
\Psi \right) \widehat{\mathbf{z}}=\mathbf{0}
\end{equation*}%
\begin{equation*}
{\Large \nabla }{\large p\ }+{\Large \nabla }^{2}\Psi {\Large \nabla }\Psi
+B_{z}{\Large \nabla }B_{z}=\mathbf{0}
\end{equation*}%
{\large which represents the '\textit{equivalent GSEs set}'},{\large \ or}%
\begin{equation}
{\Large \nabla }\Psi \times {\Large \nabla }B_{z}\cdot \widehat{\mathbf{z}}=0
\tag{2.7}  \label{2.7}
\end{equation}%
\begin{equation}
{\Large \nabla }{\large p\ }+{\Large \nabla }^{2}\Psi {\Large \nabla }\Psi
+B_{z}{\Large \nabla }B_{z}=\mathbf{0}  \tag{2.8}  \label{2.8}
\end{equation}%
{\large where the equation (2.7) derives from}%
\begin{equation}
\widehat{\mathbf{z}}\times {\Large \nabla }B_{z}\cdot {\Large \nabla }\Psi =0
\tag{2.9}  \label{2.9}
\end{equation}%
\ {\large The equations (2.7) and (2.8) are the vectorial-scalar form system
of the GSE for the ideal magnetohydrodynamical equilibria in a
two-dimensional framework and they lead to a fundamental 2nd-order PDEs
system for searching the steady-state solution of the stationary problem for
a fully ionized plasma (astrophysical plasma).}

\subsubsection{Canonical form of the GSE}

{\large We can observe that} {\large the \textit{canonical form} of the GSE
derives from the system (2.7),(2.8) as it follows: by the equation (2.7) we
have necessarily}%
\begin{equation}
{\Large \nabla }\Psi \times {\Large \nabla }B_{z}=\mathbf{0}\text{ \ \ \ }%
\forall (x,y)\in Dom(\Psi )\cap Dom(B_{z})  \tag{2.10}  \label{2.10}
\end{equation}%
{\large because }${\Large \nabla }\Psi \times {\Large \nabla }B_{z}$ {\large %
is on the }$z$ {\large axis; this one delivers a linear dependence between }$%
{\Large \nabla }\Psi $ {\large and}$\ {\Large \nabla }B_{z}$ {\large as it
follows }%
\begin{equation}
{\Large \nabla }\Psi =\alpha (\Psi ,B_{z}){\Large \nabla }B_{z}  \tag{2.11}
\label{2.11}
\end{equation}%
{\large where we suppose that }$\alpha $ {\large is a rational real
functional of }$\Psi $ {\large and }${\large B}_{z}${\large \ defined in} $%
Dom(\Psi )\cap Dom(B_{z})${\large ;} {\large \ therefore\ we obtain} {\large %
by the (2.8)}%
\begin{equation*}
{\Large \nabla }{\large p\ }+{\Large \nabla }^{2}\Psi \alpha (\Psi ,B_{z})%
{\Large \nabla }B_{z}+B_{z}{\Large \nabla }B_{z}=\mathbf{0}
\end{equation*}%
{\large which delivers}%
\begin{equation}
{\Large \nabla }{\large p+}\left( {\Large \nabla }^{2}\Psi \alpha (\Psi
,B_{z})+B_{z}\right) {\Large \nabla }B_{z}=\mathbf{0}  \tag{2.12}
\label{2.12}
\end{equation}%
{\large this one means that }${\Large \nabla }B_{z}$ {\large and} ${\Large %
\nabla }{\large p}$ {\large are linearly dipendent among them, or they are
co-axial vectors; furthermore, by applying }${\Large \nabla \Psi \times }$%
{\large \ to this one, we obtain}%
\begin{equation}
{\large \nabla \Psi \times \nabla p}=\mathbf{0}\text{ \ \ \ \ }\forall
(x,y)\in \left( Dom(\Psi )\cap Dom({\large p})\right)  \tag{2.13}
\label{2.13}
\end{equation}%
{\large which means that }${\Large \nabla \Psi }${\large \ and }${\Large %
\nabla }{\large p}${\large \ are necessarily linearly dependent as in
(2.10); thus determines}%
\begin{equation}
{\large \nabla \Psi }=\beta (\Psi ,{\large p}){\Large \nabla }{\large p} 
\tag{2.14}  \label{2.14}
\end{equation}%
{\large where we suppose that }$\beta $ {\large is a rational real
functional of }$\Psi $ {\large and }${\large p}$ {\large defined in} $%
Dom(\Psi )\cap Dom({\large p})${\large ; now, we can observe that (2.14)
lead to the differential equation}%
\begin{equation*}
\mathbf{0}={\Large \nabla }\Psi -\beta (\Psi ,{\large p}){\Large \nabla }%
{\large p}=\widehat{D}(\Psi ,{\large p})
\end{equation*}%
{\large which is}%
\begin{equation}
\mathbf{0}=\widehat{D}(\Psi ,{\large p})\text{\ \ \ }\forall (x,y)\in \left(
Dom(\Psi )\cap Dom({\large p})\right)  \tag{2.15}  \label{2.15}
\end{equation}%
{\large where we suppose that }$\widehat{D}$ {\large is a real differential
functional dependent on }$\beta $ {\large or}%
\begin{equation*}
\widehat{D}(\Psi ,{\large p})(x,y)\in 
\mathbb{R}
\text{ \ \ \ \ }\forall (x,y)\in Dom(\Psi )\cap Dom({\large p})
\end{equation*}%
{\large a general integral of (2.15) can be written as }%
\begin{equation}
{\large p}=F_{\beta }(\Psi )\ \ \ \ \forall (x,y)\in \Omega _{\beta
}\subseteq \left( Dom(\Psi )\cap Dom({\large p})\right)  \tag{2.16}
\label{2.16}
\end{equation}%
{\large where }$F_{\beta }$ {\large is a real functional dependent on }$%
\beta ${\large \ what is defined in }$\Omega _{\beta }$;

{\large If we take the equation (2.11) instead, we have a differential
equation equal to}%
\begin{equation*}
\mathbf{0}={\Large \nabla }\Psi -\alpha (\Psi ,B_{z}){\Large \nabla }B_{z}=%
\widetilde{D}(\Psi ,B_{z})
\end{equation*}%
{\large which is}%
\begin{equation}
\mathbf{0}=\widetilde{D}(\Psi ,B_{z})\text{\ \ \ }\forall (x,y)\in \left(
Dom(\Psi )\cap Dom(B_{z})\right)  \tag{2.17}  \label{2.17}
\end{equation}%
{\large where we suppose, as above, that }$\widetilde{{\large D}}$ {\large %
is a differential functional dependent on }$\alpha $; {\large a general
integral of (2.17) therefore can be written as }%
\begin{equation}
B_{z}=F_{\alpha }(\Psi )\text{ \ \ \ \ }\forall (x,y)\in \Omega _{\alpha
}\subseteq \left( Dom(\Psi )\cap Dom(B_{z})\right)  \tag{2.18}  \label{2.18}
\end{equation}%
{\large where }$F_{\alpha }$ {\large is a real functional dependent on }$%
\alpha ${\large \ and it's defined in }$\Omega _{\alpha }$;

{\large now, if }$\widehat{b}\neq \overrightarrow{\mathbf{0}}$ {\large is
the unit vector for both }${\Large \nabla }B_{z}$ {\large and }${\Large %
\nabla }{\large p}${\large , from (2.12) it follows}%
\begin{eqnarray*}
\left\Vert {\Large \nabla }{\large p}\right\Vert \widehat{b}+\left( {\Large %
\nabla }^{2}\Psi \alpha (\Psi ,B_{z})+B_{z}\right) \left\Vert {\Large \nabla 
}B_{z}\right\Vert \widehat{b} &=&\mathbf{0} \\
\left[ \left\Vert {\Large \nabla }{\large p}\right\Vert +\left( {\Large %
\nabla }^{2}\Psi \alpha (\Psi ,B_{z})+B_{z}\right) \left\Vert {\Large \nabla 
}B_{z}\right\Vert \right] \widehat{b} &=&\mathbf{0} \\
\left\Vert {\Large \nabla }{\large p}\right\Vert +\left( {\Large \nabla }%
^{2}\Psi \alpha (\Psi ,B_{z})+B_{z}\right) \left\Vert {\Large \nabla }%
B_{z}\right\Vert &=&0
\end{eqnarray*}%
{\large which lead to}%
\begin{equation}
{\Large \nabla }^{2}\Psi =\left( -\left\Vert {\Large \nabla }{\large p}%
\right\Vert \left\Vert {\Large \nabla }B_{z}\right\Vert ^{-1}-B_{z}\right)
\alpha (\Psi ,B_{z})^{-1}  \tag{2.19}  \label{2.19}
\end{equation}%
{\large and taking in account (2.16) and (2.18)\ in this one, we obtain
finally the GSE in canonical form }%
\begin{equation}
{\Large \nabla }^{2}\Psi =\left( -\left\Vert {\Large \nabla }F_{\beta }(\Psi
)\right\Vert \left\Vert {\Large \nabla }F_{\alpha }(\Psi )\right\Vert
^{-1}-F_{\alpha }(\Psi )\right) \alpha (\Psi ,F_{\alpha }(\Psi ))^{-1} 
\tag{2.20}  \label{2.20}
\end{equation}%
{\large in which we can recognize an Helmholtz Equation. In this one we
suppose that}%
\begin{equation*}
\underset{(x,y)\rightarrow (\overline{x},\overline{y})}{\lim }\frac{%
\left\Vert {\Large \nabla }F_{\beta }(\Psi )\right\Vert }{\left\Vert {\Large %
\nabla }F_{\alpha }(\Psi )\right\Vert }\in 
\mathbb{R}
\text{,\ \ \ }\forall (\overline{x},\overline{y})\in \Omega _{\alpha }\cap
\Omega _{\beta }:\left\Vert {\Large \nabla }F_{\alpha }(\Psi )\right\Vert =0
\end{equation*}%
{\large because the flux function }$\Psi ${\large \ has not any pole inside\
the equilibria domain }$\Omega _{\alpha }\cap \Omega _{\beta }$.

\subsection{A general choice in the complex plane for the free fields}

{\large A pseudo-general form of the NLSE show at first view a more great
complexity than an ordinary cubic NLSE cited in }\textit{Lapenta}{\large 
\textit{'s paper }but on the other hand its mathematical morphology is
affected by a more general set of solutions, which can represent a
wide-spectrum of possibilities for the research of explanations in the
physical framework for the above equilibrium structures. Furthermore, the
general form of the NLSE cited below includes both }\textit{Lapenta}{\large %
\ choice \cite{8} and }\textit{Throumoulopoulos et al. }{\large choice \cite%
{9} (this last leads to a pseudo-cubic NLSE); in this way, several
specialized choices set for a same physical problem can be put in a more
general mathematical solving context from the point of view of the related
non-linear problem. In this case, such a strategy therefore permits us to
define a valid vay for to find a more general possible solving method for a
GSEs set in object, working in the framework of the solitonic solutions as
in the interesting idea of }\textit{Lapenta}{\large .}

{\large A pseudo-general form of the NLSE, in the time-dependent case, can
be represented as it follows \cite{6} }%
\begin{equation}
{\large i}\frac{\partial \Psi }{\partial t}{\large =\Delta }_{3}{\large \Psi
+g}\left( \left\vert \Psi \right\vert ^{2},t\right) {\large \Psi +i\sigma }%
\left( \left\vert \Psi \right\vert ^{2},t\right) {\large \Psi }  \tag{2.21}
\label{2.21}
\end{equation}%
{\large where }${\large \Psi }$ {\large is a complex function, }${\Large %
\Delta }_{3}$ {\large is the Laplace operator in the }$(x,y,z)$ {\large space%
} {\large framework and both\ }$g$ {\large and}$\ \sigma $ {\large are
supposed to be time-dependent potential complex functionals of }$\left\vert
\Psi \right\vert ^{2}$. {\large It's remarkable the fact that this equation
is named 'pseudo-general' because the functions \ }$g$ {\large and}$\ \sigma 
$ {\large are complex-valued functionals} {\large instead that real-valued
functionals\ as is usually}. {\large However,} {\large it's clear that
(2.21) can be put 'always' in a form in which }$g$ {\large and}$\ \sigma $ 
{\large lead to a pair of real-valued functionals. Indeed}%
\begin{eqnarray*}
{\large g\Psi +i\sigma \Psi } &{\large =}&\left( {\large g}_{R}+i{\large g}%
_{i}\right) {\large \Psi +i}\left( {\large \sigma }_{R}+i{\large \sigma }%
_{i}\right) {\large \Psi =} \\
&{\large =}&\left( {\large g}_{R}-{\large \sigma }_{i}\right) {\large \Psi +i%
}\left( {\large g}_{i}+{\large \sigma }_{R}\right) {\large \Psi =}\text{ }%
\widetilde{{\large g}}{\large \Psi +i}\widetilde{{\large \sigma }}{\large %
\Psi }
\end{eqnarray*}%
{\large where }$\widetilde{{\large g}}${\large \ and\ }$\widetilde{{\large %
\sigma }}$ {\large are obviously real functionals. In the stationary case,
this equation leads to the steady-state form}%
\begin{equation}
{\large \Delta }_{3}{\large \Psi +}\left( {\large g}\left( \left\vert \Psi
\right\vert ^{2}\right) \text{ }{\large +}\text{ }{\large i\sigma }\left(
\left\vert \Psi \right\vert ^{2}\right) \right) {\large \Psi =0}  \tag{2.22}
\label{2.22}
\end{equation}%
{\large which is, as we see, a more general form of the Helmholtz Equation;
it can be viewed as \cite{7}}%
\begin{equation}
\left( {\large \Delta }_{3}{\large +\varpi }^{2}\left( \left\vert \Psi
\right\vert ^{2}\right) \right) {\large \Psi =0}  \tag{2.23}  \label{2.23}
\end{equation}%
{\large where }%
\begin{equation}
{\large \varpi }^{2}\left( \left\vert \Psi \right\vert ^{2}\right) ={\large g%
}\left( \left\vert \Psi \right\vert ^{2}\right) {\large +i\sigma }\left(
\left\vert \Psi \right\vert ^{2}\right)  \tag{2.24}  \label{2.24}
\end{equation}%
{\large is a complex-valued function. This non-linear PDE therefore
represents the fundamental link between the GSE and the pseudo-general NLSE
in the stationary case for a three-dimensional equilibrium problem. In the
two-dimensional framework }$(x,y)$, {\large that is the framework in which
we analize the localized accretion plasma structures,} {\large we have
therefore the differential equation}%
\begin{equation}
\left( {\large \Delta }_{2}{\large +\varpi }^{2}\left( \left\vert \Psi
\right\vert ^{2}\right) \right) {\large \Psi =0}\text{ \ \ {\large that is \
\ }}{\Large \nabla }_{{\small 2}}^{2}{\large \Psi =-\varpi }^{2}\left(
\left\vert \Psi \right\vert ^{2}\right) {\large \Psi }  \tag{2.25}
\label{2.25}
\end{equation}%
{\large At this point, we observe that an appropriate choice for the free
fields in the GSEs can permit the translating of such equilibrium equations
in a certain form of the NLSE, as the general stationary form (2.22), which
could be an excellent solution for re-interpretate localized structures in
the accretion discs as 'solitonic structures', going as viewed by the
Helmholtz Equation. Indeed we'll see that a general transformations set for
the free fields }${\Large p}$ {\large and} ${\Large B}_{z}${\large \ can
lead from the equations (2.7) and (2.8) to the equation (2.25) by an
appropriate expression for the }${\large \varpi }^{2}$ {\large function on
the same }${\Large p}$ {\large and} ${\Large B}_{z}$. {\large It's important
to observe that the 'analiticity' of the complex function }$\Psi $ {\large %
is not required for solving the Helmholtz Equation in general \cite{7}.}

{\large Furthermore}, {\large I cite that the 'moment method' \cite{6} can
be used as a valid approximation analytical way for to solve a wide family
of non-linear wave equations of NLSE type in its pseudo-general form; this
method have been developed for n-dimensional cases in general by }\textit{%
Garc\'{\i}a-Ripoll, J.J.}{\large \ and }\textit{P\'{e}rez-Garc\'{\i}a, V.M.}%
{\large . So, it's remarkable that by this mathematical approximation
strategy it could be possible to identify a good set of physically coherent
solutions of the GSE for the magnetohydrodynamical equilibrium problem, i.e.
solitonic solutions as we would like.}

\subsubsection{Connection between GSEs and a pseudo-general form of the NLSE}

{\large Now}, {\large we consider the expressions (2.16) and (2.18) in the
intersection of the }$\Psi ,{\large B}_{z}${\large \ and }${\large p}$%
{\large \ domains}%
\begin{equation*}
{\large p}=F_{\beta }(\Psi )\text{ \ \ \ \ }B_{z}=F_{\alpha }(\Psi )\text{ \
\ \ \ \ }\forall (x,y)\in \Omega _{\alpha }\cap \Omega _{\beta }
\end{equation*}%
{\large and replacing the free fields }${\large p}$ {\large and} ${\large B}%
_{z}$ {\large in the equation (2.8) with this relations we obtain}%
\begin{equation*}
{\Large \nabla }F_{\beta }(\Psi ){\large \ }+{\Large \nabla }^{2}\Psi 
{\Large \nabla }\Psi +F_{\alpha }(\Psi ){\Large \nabla }F_{\alpha }(\Psi )=%
\mathbf{0}
\end{equation*}%
{\large obviously for the linear dependence of }${\Large \nabla }{\large p\ }%
+B_{z}{\Large \nabla }B_{z}$ {\large by }${\Large \nabla }^{2}\Psi {\Large %
\nabla }\Psi $\ {\large this one delivers}%
\begin{equation}
{\Large \nabla }F_{\beta }(\Psi ){\large \ }+F_{\alpha }(\Psi ){\Large %
\nabla }F_{\alpha }(\Psi )={\Large k}\left( \Psi \right) {\Large \nabla }%
\Psi =-{\Large \nabla }^{2}\Psi {\Large \nabla }\Psi  \tag{2.26}
\label{2.26}
\end{equation}%
{\large or}%
\begin{eqnarray}
\left( {\Large \nabla }^{2}\Psi +{\Large k}\left( \Psi \right) \right) 
{\Large \nabla }\Psi &=&\mathbf{0}\text{ \ \ \ \ }  \TCItag{2.27}
\label{2.27} \\
\text{{\large which is} \ \ \ \ }{\Large \nabla }^{2}\Psi +{\Large k}\left(
\Psi \right) &=&0\text{ \ \ \ }\forall (x,y)\in \Omega _{\alpha }\cap \Omega
_{\beta }  \notag
\end{eqnarray}%
{\large that is exactly the Helmholtz Equation, in which we suppose that }$%
{\large k}$ {\large is a generic functional of }$\Psi ${\large ; if we
compare the second equation in (2.27) with the Helmholtz Equation (2.25) and
we consider the expression (2.24), then we obtain necessarily}%
\begin{equation}
{\Large k}\left( \Psi \right) ={\large \varpi }^{2}\left( \left\vert \Psi
\right\vert ^{2}\right) {\large \Psi }=\left( {\large g}\left( \left\vert
\Psi \right\vert ^{2}\right) {\large +i\sigma }\left( \left\vert \Psi
\right\vert ^{2}\right) \right) {\large \Psi }\text{ \ \ \ }\forall (x,y)\in
\Omega _{\alpha }\cap \Omega _{\beta }  \tag{2.28}  \label{2.28}
\end{equation}

{\large which indicates that }${\Large k}${\large \ is a complex functional
and\ }$\Psi $ {\large is clearly a complex function by the equation (2.27),
because by this one it derives}%
\begin{eqnarray}
{\Large \nabla }^{2}\Psi &=&-{\Large k}\left( \Psi \right) =-\left( {\large g%
}\left( \left\vert \Psi \right\vert ^{2}\right) {\large +i\sigma }\left(
\left\vert \Psi \right\vert ^{2}\right) \right) \Psi  \notag \\
\text{{\large or} \ \ \ \ \ }\Psi ^{-1}{\Large \nabla }^{2}\Psi &=&-\left( 
{\large g}\left( \left\vert \Psi \right\vert ^{2}\right) {\large +i\sigma }%
\left( \left\vert \Psi \right\vert ^{2}\right) \right) \text{ \ \ \ {\large %
with\ \ }}{\large g},{\large \sigma \neq }\text{ }0  \TCItag{2.29}
\label{2.29}
\end{eqnarray}%
{\large \ At this point, taking into account the }\textit{Lapenta}{\large \
and }\textit{Throumoulopoulos et al.}{\large 's works \cite{8}, \cite{9} and
starting from their fields choices, I found a well posed pair of
transformation functionals like this}%
\begin{equation}
{\large B}_{z}{\large \nabla B}_{z}=\frac{1}{2}{\large \nabla B}_{z}^{2}=%
\underset{i,j}{\sum }a_{ij}\overline{\Psi }^{i}\Psi ^{j+1}{\Large \nabla }%
\Psi  \tag{2.30}  \label{2.30}
\end{equation}%
\begin{equation}
{\large \nabla p}=\underset{i,j}{\sum }b_{ij}\overline{\Psi }^{i}\Psi ^{j+1}%
{\Large \nabla }\Psi  \tag{2.31}  \label{2.31}
\end{equation}%
{\large where }$\overline{\Psi }$ {\large is the complex conjugate function
related to the complex flux function }$\Psi $ {\large and }$a_{ij},b_{ij}$ 
{\large are complex constants}; {\large we observe that (2.30) and (2.31)
are respectively a term proportional to the magnetic force density and the
pure hydrodynamical pressure force density, which are two 'real' physical
variables; this is for me a fundamental physical condition and it's
important that it doesn't affect the complexity of the flux function }$\Psi $%
; {\large we} {\large will explore this 'reality condition' in the next
section as a 'validity condition' for (2.30) and (2.31). From these
transformations and the first equation in (2.26) it derives}%
\begin{equation}
{\Large \nabla }{\large p\ }+B_{z}{\Large \nabla }B_{z}=\underset{i,j}{\sum }%
\left( a_{ij}+b_{ij}\right) \overline{\Psi }^{i}\Psi ^{j+1}{\Large \nabla }%
\Psi ={\Large k}\left( \Psi \right) {\Large \nabla }\Psi  \tag{2.32}
\label{2.32}
\end{equation}%
{\large which together with (2.28) delivers}%
\begin{eqnarray}
{\Large k}\left( \Psi \right) &=&\underset{i,j}{\sum }\left(
a_{ij}+b_{ij}\right) \overline{\Psi }^{i}\Psi ^{j+1}=  \TCItag{2.33}
\label{2.33} \\
&=&\left( {\large g}\left( \left\vert \Psi \right\vert ^{2}\right) {\large %
+i\sigma }\left( \left\vert \Psi \right\vert ^{2}\right) \right) \Psi \text{
\ \ \ }\forall (x,y)\in \Omega _{\alpha }\cap \Omega _{\beta }  \notag
\end{eqnarray}%
{\large Now, it's interesting to observe that} {\large the fields
transformations (2.30), (2.31) lead from equations (2.7),(2.8) to the
pseudo-general form (2.25) of the NLSE in a space framework }$(x,y)$ {\large %
by the equation}%
\begin{eqnarray}
\underset{i,j}{\sum }\gamma _{ij}\overline{\Psi }^{i}\Psi ^{j} &=&{\large g}%
\left( \left\vert \Psi \right\vert ^{2}\right) {\large +i\sigma }\left(
\left\vert \Psi \right\vert ^{2}\right) \text{ \ \ \ }\forall (x,y)\in
\Omega _{\alpha }\cap \Omega _{\beta }  \TCItag{2.34}  \label{2.34} \\
\text{{\large where} \ \ \ \ \ }\gamma _{ij} &=&a_{ij}+b_{ij}\in 
\mathbb{C}
\text{ \ \ {\large and}\ \ }\func{Im}\left( {\large \Psi }\right) \subset 
\mathbb{C}
\notag
\end{eqnarray}%
{\large Because of the second equation in (2.29), we will proof below that,
if }$\Psi $ {\large is a complex function, }$\underset{i,j}{\sum }\gamma
_{ij}\overline{\Psi }^{i}\Psi ^{j}${\large \ can be put in the form}%
\begin{equation*}
{\large g}\left( \left\vert \Psi \right\vert ^{2}\right) {\large +i\sigma }%
\left( \left\vert \Psi \right\vert ^{2}\right)
\end{equation*}%
{\large where both\ }$g$ {\large and}$\ \sigma $ {\large are supposed to be
complex functionals as above.}

{\large In this regard}, {\large we consider the expression at left hand
side of (2.34) in }$\Omega _{\alpha }\cap \Omega _{\beta }$ {\large domain}%
\begin{equation}
\underset{i,j}{\sum }\gamma _{ij}\overline{\Psi }^{i}\Psi ^{j}  \tag{2.35}
\label{2.35}
\end{equation}

{\large and if we develop the sum, we obtain the next identities chain}%
\begin{eqnarray*}
\underset{i,j}{\sum }\gamma _{ij}\overline{\Psi }^{i}\Psi ^{j} &=&\underset{j%
}{\sum }\gamma _{0j}\Psi ^{j}+\underset{j}{\sum }\gamma _{1j}\overline{\Psi }%
\Psi ^{j}+\underset{j}{\sum }\gamma _{2j}\overline{\Psi }^{2}\Psi ^{j}+...=
\\
&=&\underset{j}{\sum }\gamma _{0j}\Psi ^{j}+\underset{j=1}{\overset{...}{%
\sum }}\gamma _{1j}\left\vert \Psi \right\vert ^{2}\Psi ^{j-1}+\underset{j=2}%
{\overset{...}{\sum }}\gamma _{2j}\left( \left\vert \Psi \right\vert
^{2}\right) ^{2}\Psi ^{j-2}+...=
\end{eqnarray*}%
\begin{equation*}
=\left( \gamma _{00}+\gamma _{10}\left\vert \Psi \right\vert ^{2}+\gamma
_{20}\left( \left\vert \Psi \right\vert ^{2}\right) ^{2}+...\right) +
\end{equation*}%
\begin{equation*}
+\left( \gamma _{01}+\gamma _{11}\left\vert \Psi \right\vert ^{2}+\gamma
_{21}\left( \left\vert \Psi \right\vert ^{2}\right) ^{2}+...\right) \Psi
+...=
\end{equation*}%
\begin{eqnarray*}
&=&\underset{i}{\sum }\gamma _{i0}\left( \left\vert \Psi \right\vert
^{2i}\right) \Psi ^{0}+\underset{i}{\sum }\gamma _{i1}\left( \left\vert \Psi
\right\vert ^{2i}\right) \Psi ^{1}+\underset{i}{\sum }\gamma _{i2}\left(
\left\vert \Psi \right\vert ^{2i}\right) \Psi ^{2}+...= \\
&=&\underset{j}{\sum }\left( \underset{i}{\sum }\gamma _{ij}\left(
\left\vert \Psi \right\vert ^{2i}\right) \right) \Psi ^{j}=\underset{j}{\sum 
}f_{j}\left( \left\vert \Psi \right\vert ^{2}\right) \Psi ^{j}
\end{eqnarray*}%
{\large which deliver}%
\begin{equation}
\underset{i,j}{\sum }\gamma _{ij}\overline{\Psi }^{i}\Psi ^{j}=\underset{j}{%
\sum }f_{j}\left( \left\vert \Psi \right\vert ^{2}\right) \Psi ^{j} 
\tag{2.36}  \label{2.36}
\end{equation}%
{\large where}%
\begin{equation}
f_{j}\left( \left\vert \Psi \right\vert ^{2}\right) =\underset{i}{\sum }%
\gamma _{ij}\left( \left\vert \Psi \right\vert ^{2i}\right)   \tag{2.37}
\label{2.37}
\end{equation}%
{\large is a complex polynomial functional on }$\left\vert \Psi \right\vert
^{2i}$. {\large Now, we 'impose' the complexity of the flux function }$\Psi $
($\func{Im}\left( {\large \Psi }\right) \subset 
\mathbb{C}
$) {\large and we write therefore}%
\begin{equation}
\Psi (x,y)=u(x,y)+iv(x,y)\text{ \ \ \ }\forall (x,y)\in \Omega _{\alpha
}\cap \Omega _{\beta }  \tag{2.38}  \label{2.38}
\end{equation}%
{\large where }$u$ {\large and}\ $v$ {\large are generic real-valued
functions in the two-dimensional framework} $(x,y)$; {\large from this one
and (2.36) we obtain}%
\begin{eqnarray}
\underset{i,j}{\sum }\gamma _{ij}\overline{\Psi }^{i}\Psi ^{j}\underset{j}{%
=\sum }f_{j}\left( \left\vert \Psi \right\vert ^{2}\right) \left(
u+iv\right) ^{j} &=&  \TCItag{2.39}  \label{2.39} \\
\underset{j}{=\sum }f_{j}\left( \left\vert \Psi \right\vert ^{2}\right) 
\underset{k=0}{\overset{j}{\sum }}\binom{j}{k}u^{j-k}(iv)^{k} &=&\underset{j}%
{\text{ }\sum }f_{j}\left( \left\vert \Psi \right\vert ^{2}\right) \underset{%
k=0}{\overset{j}{\sum }}\zeta _{jk}u^{j-k}(iv)^{k}  \notag
\end{eqnarray}%
{\large in which we used the 'Newton binomial formula' and }$\zeta _{jk}=%
\binom{j}{k}$ {\large are the binomial coefficients.}

{\large At this point, we fix the maximum level }$N$ {\large of the
polynomial sum in (2.39) and we observe that}%
\begin{eqnarray}
\underset{j=0}{\text{ }\overset{N}{\sum }}f_{j}\left( \left\vert \Psi
\right\vert ^{2}\right) \underset{k=0}{\overset{j}{\sum }}\zeta
_{jk}u^{j-k}(iv)^{k} &=&\underset{j=0}{\text{ }\overset{N}{\sum }}\underset{%
k=0}{\overset{j}{\sum }}f_{j}\left( \left\vert \Psi \right\vert ^{2}\right)
\zeta _{jk}u^{j-k}(iv)^{k}=  \notag \\
&=&\underset{j=0}{\text{ }\overset{N}{\sum }}\underset{k=0}{\overset{j}{\sum 
}}\eta _{jk}\left( \left\vert \Psi \right\vert ^{2}\right) u^{j-k}(iv)^{k} 
\TCItag{2.40}  \label{2.40}
\end{eqnarray}

\begin{equation}
\text{{\large where \ \ \ \ \ }}\eta _{jk}\left( \left\vert \Psi \right\vert
^{2}\right) =f_{j}\left( \left\vert \Psi \right\vert ^{2}\right) \zeta _{jk}=%
\binom{j}{k}\underset{i}{\sum }\gamma _{ij}\left( \left\vert \Psi
\right\vert ^{2i}\right)  \tag{2.41}  \label{2.41}
\end{equation}

{\large is a new complex polynomial functional; from (2.40) therefore it
derives}%
\begin{equation*}
\underset{j=0}{\text{ }\overset{N}{\sum }}\underset{k=0}{\overset{j}{\sum }}%
\eta _{jk}\left( \left\vert \Psi \right\vert ^{2}\right) u^{j-k}(iv)^{k}=
\end{equation*}%
\begin{equation*}
=\text{ }\underset{l=0}{\overset{N}{\sum }}\eta _{l0}\left( \left\vert \Psi
\right\vert ^{2}\right) u^{l}(iv)^{0}+\underset{l=0}{\overset{N-1}{\sum }}%
\eta _{\left( l+1\right) 1}\left( \left\vert \Psi \right\vert ^{2}\right)
u^{l}(iv)^{1}+\underset{l=0}{\overset{N-2}{\sum }}\eta _{\left( l+2\right)
2}\left( \left\vert \Psi \right\vert ^{2}\right) u^{l}(iv)^{2}+...
\end{equation*}

{\large that is}%
\begin{equation*}
\underset{h=0}{\text{ }\overset{N/2}{\sum }}\left( \underset{l=0}{\overset{%
N-2h}{\sum }}\eta _{l\left( 2h\right) }\left( \left\vert \Psi \right\vert
^{2}\right) u^{l}\right) (iv)^{2h}+\underset{h=0}{\text{ }\overset{N/2-1}{%
\sum }}\left( \underset{l=0}{\overset{N-\left( 2h+1\right) }{\sum }}\eta
_{l\left( 2h+1\right) }\left( \left\vert \Psi \right\vert ^{2}\right)
u^{l}\right) (iv)^{2h+1}
\end{equation*}%
{\large if }$N${\large \ is 'even'; in clear complex form this one deliver,}%
\begin{equation}
\underset{i,j}{\sum }\gamma _{ij}\overline{\Psi }^{i}\Psi ^{j}=  \tag{2.42}
\label{2.42}
\end{equation}%
\begin{eqnarray*}
&=&\underset{h=0}{\text{ }\overset{\frac{N}{2}}{\sum }}\left( \underset{l=0}{%
\overset{N-2h}{\sum }}\eta _{l\left( 2h\right) }\left( \left\vert \Psi
\right\vert ^{2}\right) u^{l}\right) \left( -1\right) ^{h}v^{2h}+ \\
&&+\text{ }{\Large i}\underset{h=0}{\overset{\frac{N}{2}-1}{\sum }}\left( 
\underset{l=0}{\overset{N-\left( 2h+1\right) }{\sum }}\eta _{l\left(
2h+1\right) }\left( \left\vert \Psi \right\vert ^{2}\right) u^{l}\right)
\left( -1\right) ^{h}v^{2h+1}
\end{eqnarray*}%
{\large and if it's compared with equation (2.34), taking into account
(2.41), it delivers finally the next expressions for the }$g$ {\large and}$\
\sigma $ {\large functionals }

$\forall (x,y)\in \Omega _{\alpha }\cap \Omega _{\beta }${\large \ \ \ \ }%
\begin{equation*}
{\large g}\left( \left\vert \Psi \right\vert ^{2},u,v\right) =\underset{h=0}{%
\text{ }\overset{\frac{N}{2}}{\sum }}\left( \underset{l=0}{\overset{N-2h}{%
\sum }}\eta _{l\left( 2h\right) }\left( \left\vert \Psi \right\vert
^{2}\right) u^{l}\right) \left( -1\right) ^{h}v^{2h}=
\end{equation*}%
\begin{equation}
=\underset{i}{\sum }\underset{h=0}{\overset{\frac{N}{2}}{\sum }}\underset{l=0%
}{\overset{N-2h}{\sum }}\binom{l}{2h}\gamma _{il}\left( \left\vert \Psi
\right\vert ^{2i}\right) u^{l}\left( -1\right) ^{h}v^{2h}  \tag{2.43a}
\label{2.43a}
\end{equation}

\begin{equation*}
{\large \sigma }\left( \left\vert \Psi \right\vert ^{2},u,v\right) =\underset%
{h=0}{\overset{\frac{N}{2}-1}{\sum }}\left( \underset{l=0}{\overset{N-\left(
2h+1\right) }{\sum }}\eta _{l\left( 2h+1\right) }\left( \left\vert \Psi
\right\vert ^{2}\right) u^{l}\right) \left( -1\right) ^{h}v^{2h+1}=
\end{equation*}%
\begin{equation}
=\underset{i}{\sum }\underset{h=0}{\overset{\frac{N}{2}}{\sum }}\underset{l=0%
}{\overset{N-\left( 2h+1\right) }{\sum }}\binom{l}{2h+1}\gamma _{il}\left(
\left\vert \Psi \right\vert ^{2i}\right) u^{l}\left( -1\right) ^{h}v^{2h+1} 
\tag{2.44a}  \label{2.44a}
\end{equation}
{\large which are clearly two complex functionals on the square function }$%
\left\vert \Psi \right\vert ^{2}=\overline{\Psi }\Psi $ {\large and on the }$%
u$ {\large and} $v$ {\large functions; they are complex because the presence
of the complex coefficients }$\gamma _{il}${\large \ in the above
expressions.}

{\large If the level of the polynomial sum }$N$ {\large is odd, we have
similarly the complex functionals}%
\begin{equation}
{\large g}\left( \left\vert \Psi \right\vert ^{2},u,v\right) =\underset{h=0}{%
\text{ }\overset{\frac{N-1}{2}}{\sum }}\left( \underset{l=0}{\overset{N-2h}{%
\sum }}\eta _{l\left( 2h\right) }\left( \left\vert \Psi \right\vert
^{2}\right) u^{l}\right) \left( -1\right) ^{h}v^{2h}=  \notag
\end{equation}%
\begin{equation}
=\underset{i}{\sum }\underset{h=0}{\text{ }\overset{\frac{N-1}{2}}{\sum }}%
\underset{l=0}{\overset{N-2h}{\sum }}\binom{l}{2h}\gamma _{il}\left(
\left\vert \Psi \right\vert ^{2i}\right) u^{l}\left( -1\right) ^{h}v^{2h} 
\tag{2.43b}  \label{2.43b}
\end{equation}

\begin{equation}
{\large \sigma }\left( \left\vert \Psi \right\vert ^{2},u,v\right) =\underset%
{h=0}{\overset{\frac{N-1}{2}}{\sum }}\left( \underset{l=0}{\overset{N-\left(
2h+1\right) }{\sum }}\eta _{l\left( 2h+1\right) }\left( \left\vert \Psi
\right\vert ^{2}\right) u^{l}\right) \left( -1\right) ^{h}v^{2h+1}=  \notag
\end{equation}%
\begin{equation}
=\underset{i}{\sum }\underset{h=0}{\text{ }\overset{\frac{N-1}{2}}{\sum }}%
\underset{l=0}{\overset{N-\left( 2h+1\right) }{\sum }}\binom{l}{2h+1}\gamma
_{il}\left( \left\vert \Psi \right\vert ^{2i}\right) u^{l}\left( -1\right)
^{h}v^{2h+1}  \tag{2.44b}  \label{2.44b}
\end{equation}
{\large We observe that in these expressions the maximum level of polynomial
sum on }$i${\large -index} {\large is not fixed, leaving in this way the
choice of dependence of the above functionals on the even powers of the
function }$\left\vert \Psi \right\vert ^{2}$ {\large completely free. Such a
dependence can be imposed on the base of mathematical or physical criteria
which are well defined if they are related to a specific physical problem.}

\begin{notation}
{\large it's clear that the general choice (2.30),(2.31) includes both }%
\textit{Lapenta}{\large \ \cite{8} and }\textit{Throumoulopoulos et al. }%
{\large \cite{9} choices; indeed for the first choice we have}%
\begin{equation}
a_{00}=\alpha _{0}^{2},\text{ \ \ \ }a_{ij}=0\text{ \ \ \ }\forall i,j>0 
\tag{2.45a}  \label{2.45a}
\end{equation}
\begin{equation}
b_{11}=\alpha _{0}^{2},\text{ \ \ \ }b_{ij}=0\text{ \ \ \ }\forall \left(
i,j\right) \neq \left( 1,1\right)  \tag{2.45b}  \label{2.45b}
\end{equation}%
{\large while for the second choice we have}%
\begin{equation}
a_{00}=\alpha _{0}^{2},\text{ \ \ \ }a_{ij}=0\text{ \ \ \ }\forall i,j>0 
\tag{2.46a}  \label{2.46a}
\end{equation}
\begin{equation}
b_{03}=\alpha _{0}^{2},\text{ \ \ \ }b_{ij}=0\text{ \ \ }\forall \left(
i,j\right) \neq \left( 0,3\right)  \tag{2.46b}  \label{2.46b}
\end{equation}
\end{notation}

\section{Validity conditions in the mathematical and physical frameworks}

\bigskip {\large In this section we explore the validity conditions for the
general choice (2.30),(2.31) from the physical and mathematical points of
view. We observe that these conditions are named '\textit{critical conditions%
}' in the present paper because they are 'necessary' conditions for the
acceptability, or internal coherence, of the proposed mathematical models,
which are analized in respect of a related reasonable link between the
mathematical framework and the physical one. Furthermore, the analysis of
these theoretical positions will take into account necessary and sufficient
conditions too for the validity of unique solutions for the
magnetohydrodynamical equilibrium problem proposed.}

\subsection{On a critical condition for the general choice: mathematical
aspects}

{\large We consider the general choice for the free fields }${\large p}$%
{\large \ and }${\large B}_{z}${\large \ in the GSEs set (2.7),(2.8)}%
\begin{equation}
{\large B}_{z}{\large \nabla B}_{z}=\frac{1}{2}{\large \nabla B}_{z}^{2}=%
\underset{i,j}{\sum }a_{ij}\overline{\Psi }^{i}\Psi ^{j+1}{\Large \nabla }%
\Psi  \tag{3.1}  \label{3.1}
\end{equation}%
\begin{equation}
{\large \nabla p}=\underset{i,j}{\sum }b_{ij}\overline{\Psi }^{i}\Psi ^{j+1}%
{\Large \nabla }\Psi  \tag{3.2}  \label{3.2}
\end{equation}%
\begin{equation*}
\text{{\large with \ \ \ }}a_{ij},b_{ij}\in 
\mathbb{C}%
\end{equation*}%
{\large and for to establish a validity condition for the internal
mathematical coherence of these transformation functionals, we apply the
operator }${\large \nabla \times }$ {\large at the both sides of
(3.1),(3.2), as in the }\textit{Throumoulopoulos et al. }{\large work \cite%
{9}, taking into account that they are 'similar' except for the coefficients
sets }$\left\{ {\large a}_{ij}\right\} _{ij\in \text{ }%
\mathbb{N}
}$ {\large and} $\left\{ {\large b}_{ij}\right\} _{ij\in \text{ }%
\mathbb{N}
}${\large ; for this reason, it will be sufficient therefore to consider for
both relations the term}%
\begin{equation}
\underset{i,j}{\sum }c_{ij}\overline{\Psi }^{i}\Psi ^{j+1}{\Large \nabla }%
\Psi  \tag{3.3}  \label{3.3}
\end{equation}%
{\large with }$c_{ij}=a_{ij}$ {\large or} $b_{ij}$ {\large as appropriate,
and to reduce in normal form the expression}%
\begin{equation}
{\large \nabla \times }\underset{i,j}{\sum }c_{ij}\overline{\Psi }^{i}\Psi
^{j+1}{\Large \nabla }\Psi  \tag{3.4}  \label{3.4}
\end{equation}%
{\large for to identify a validity condition which is unique for both field
transformations. At this point the calculus is quite simple: starting from
the expression (3.4) we obtain for (3.1) and (3.2)}%
\begin{equation}
{\large \nabla \times }\underset{i,j}{\sum }c_{ij}\overline{\Psi }^{i}\Psi
^{j+1}{\Large \nabla }\Psi =\mathbf{0}\text{ \ \ }\forall (x,y)\in \Omega
_{\alpha }\cap \Omega _{\beta }  \tag{3.5}  \label{3.5}
\end{equation}%
{\large since }%
\begin{equation*}
{\large \nabla \times \nabla p=\nabla \times }\left( \frac{1}{2}{\large %
\nabla B}_{z}^{2}\right) \text{ }{\large =}\text{ }\mathbf{0}
\end{equation*}%
{\large for the 'scalar' functions }${\large p}${\large \ and }${\large B}%
_{z}^{2}$. {\large If we proceed in the calculus we obtain}%
\begin{equation*}
\mathbf{0=}\text{ }{\large \nabla \times }\underset{i,j}{\sum }c_{ij}%
\overline{\Psi }^{i}\Psi ^{j+1}{\Large \nabla }\Psi =\underset{i,j}{\sum }%
c_{ij}{\large \nabla \times }\left( \overline{\Psi }^{i}\Psi ^{j+1}{\large %
\nabla }\Psi \right)
\end{equation*}%
{\large for the linearity of vectorial product; taking in account the
vectorial identity}%
\begin{equation*}
{\large \nabla \times }\left( \alpha \overrightarrow{a}\right) \text{ }%
{\large =\alpha \nabla \times }\overrightarrow{a}\text{ }{\large -}\text{ }%
\overrightarrow{a}{\large \times \nabla \alpha }\text{ \ \ \ with{\large \ }}%
\alpha \text{ as a scalar variable,}
\end{equation*}%
{\large we have therefore (we remember that }$\Psi ${\large \ is a scalar
function)}%
\begin{eqnarray*}
\mathbf{0} &\mathbf{=}&\text{ }\underset{i,j}{\sum }c_{ij}{\large \nabla
\times }\left( \overline{\Psi }^{i}\Psi ^{j+1}{\large \nabla }\Psi \right) =%
\underset{i,j}{\sum }c_{ij}\left( \overline{\Psi }^{i}\Psi ^{j+1}{\large %
\nabla \times \nabla }\Psi -{\large \nabla }\Psi \times {\large \nabla }%
\left( \overline{\Psi }^{i}\Psi ^{j+1}\right) \right) = \\
&=&-\underset{i,j}{\text{ }\sum }c_{ij}{\large \nabla }\Psi \times {\large %
\nabla }\left( \overline{\Psi }^{i}\Psi ^{j+1}\right) =-\underset{i,j}{\text{
}\sum }c_{ij}{\large \nabla }\Psi \times \left( \Psi ^{j+1}{\large \nabla }%
\overline{\Psi }^{i}+\overline{\Psi }^{i}{\large \nabla }\Psi ^{j+1}\right) =
\\
&=&-\underset{i,j}{\text{ }\sum }c_{ij}{\large \nabla }\Psi \times \left(
\Psi ^{j+1}i\overline{\Psi }^{i-1}{\large \nabla }\overline{\Psi }+\overline{%
\Psi }^{i}(j+1)\Psi ^{j}{\large \nabla }\Psi \right) = \\
&=&-\underset{i,j}{\text{ }\sum }c_{ij}\left( \Psi ^{j+1}i\overline{\Psi }%
^{i-1}{\large \nabla }\Psi \times {\large \nabla }\overline{\Psi }+\overline{%
\Psi }^{i}(j+1)\Psi ^{j}{\large \nabla }\Psi \times {\large \nabla }\Psi
\right) = \\
&=&-\underset{i,j}{\text{ }\sum }c_{ij}\Psi ^{j+1}i\overline{\Psi }^{i-1}%
{\large \nabla }\Psi \times {\large \nabla }\overline{\Psi }=-\text{ }%
J\left( \Psi ,\left\vert \Psi \right\vert ^{2}\right) {\large \nabla }\Psi
\times {\large \nabla }\overline{\Psi }
\end{eqnarray*}%
{\large or}%
\begin{equation}
J\left( \Psi ,\left\vert \Psi \right\vert ^{2}\right) {\large \nabla }\Psi
\times {\large \nabla }\overline{\Psi }=\mathbf{0}\text{ \ \ }\forall
(x,y)\in \Omega _{\alpha }\cap \Omega _{\beta }  \tag{3.6}  \label{3.6}
\end{equation}%
{\large where}%
\begin{equation}
J\left( \Psi ,\left\vert \Psi \right\vert ^{2}\right) =\underset{i,j}{\text{ 
}\sum }c_{ij}\Psi ^{j+1}i\overline{\Psi }^{i-1}  \tag{3.7}  \label{3.7}
\end{equation}%
{\large is a complex polynomial functional of finite degree on }$\Psi $ 
{\large and}$\ \left\vert \Psi \right\vert ^{2}$. {\large As we can see,} 
{\large the equation (3.6) has the solutions}%
\begin{equation}
J\left( \Psi ,\left\vert \Psi \right\vert ^{2}\right) =0  \tag{3.8a}
\label{3.8a}
\end{equation}%
\begin{equation}
{\large \nabla }\Psi \times {\large \nabla }\overline{\Psi }=\mathbf{0} 
\tag{3.8b}  \label{3.8b}
\end{equation}%
{\large The (3.8a) solution functional equation has no solutions for an
'arbitrary' flux function }$\Psi $, {\large because the terms }$\overline{%
\Psi }^{i}\Psi ^{j}$ {\large inside it are linearly independent, or}%
\begin{equation*}
\forall i\neq \widetilde{i},\forall j\neq \widetilde{j}\text{ \ \ }\nexists 
\text{ }\omega \in 
\mathbb{C}
\setminus \left\{ 0\right\} :\overline{\Psi }^{i}\Psi ^{j}=\omega \overline{%
\Psi }^{\widetilde{i}}\Psi ^{\widetilde{j}}
\end{equation*}%
{\large indeed, for this reason it derives, taking in account (3.7),}%
\begin{equation*}
\forall \Psi ,\text{ \ }\nexists \left\{ c_{ij}\right\} _{i,j\in \text{ }%
\mathbb{N}
},\text{ }c_{ij}\in 
\mathbb{C}
:J\left( \Psi ,\left\vert \Psi \right\vert ^{2}\right) =0
\end{equation*}%
{\large this means that the functional equation (3.8a) cannot represent a
necessary condition so the equation (3.5) is verified; at this point it's
clear that the unique 'critical condition' for the mathematical coherence of
the general fields choice for an arbitrary flux function }$\Psi ${\large \ is%
}%
\begin{equation*}
{\large \nabla }\Psi \times {\large \nabla }\overline{\Psi }=\mathbf{0}\text{
\ }\forall (x,y)\in \Omega _{\alpha }\cap \Omega _{\beta }\text{\ }
\end{equation*}%
{\large Now, if we impose the complexity of this function }($\func{Im}\left( 
{\large \Psi }\right) \subset 
\mathbb{C}
$) {\large and we write therefore (see position (2.38))}%
\begin{equation*}
\Psi (x,y)=u(x,y)+iv(x,y)\text{ \ \ \ }\forall (x,y)\in \Omega _{\alpha
}\cap \Omega _{\beta }
\end{equation*}%
{\large we obtain, starting from the equation (3.8b),}%
\begin{eqnarray*}
\mathbf{0}_{%
\mathbb{C}
} &\mathbf{=}&\text{ }{\large \nabla }\Psi \times {\large \nabla }\overline{%
\Psi }={\large \nabla }\left( u+iv\right) \times {\large \nabla }\left(
u-iv\right) =\left( {\large \nabla }u+i{\large \nabla }v\right) \times
\left( {\large \nabla }u-i{\large \nabla }v\right) = \\
&=&-i{\large \nabla }u\times {\large \nabla }v+i{\large \nabla }v\times 
{\large \nabla }u=-2i{\large \nabla }u\times {\large \nabla }v
\end{eqnarray*}%
{\large (where }$\mathbf{0}_{%
\mathbb{C}
}=\mathbf{0+}i\mathbf{0}${\large ); in this way, the condition (3.8b) is
equivalent to the condition }${\large \nabla }u\times {\large \nabla }v=%
\mathbf{0}$, {\large or }%
\begin{equation*}
{\large \nabla }\Psi \times {\large \nabla }\overline{\Psi }=\mathbf{0}_{%
\mathbb{C}
}\mathbf{\Leftrightarrow }\text{ }{\large \nabla }u\times {\large \nabla }v=%
\mathbf{0}\text{ \ \ }\forall (x,y)\in \Omega _{\alpha }\cap \Omega _{\beta }
\end{equation*}%
{\large We have therefore the 'critical condition' (which is a 'necessary
condition') for both real and imaginary parts of }$\Psi $%
\begin{equation}
{\large \nabla }u\times {\large \nabla }v=\mathbf{0}\text{ \ \ }\forall
(x,y)\in \Omega _{\alpha }\cap \Omega _{\beta }  \tag{3.9}  \label{3.9}
\end{equation}%
{\large so the equation (3.6) is verified; an expansion in a two-dimensional
framework }$(x,y)${\large \ of the }${\large \nabla }$ {\large operator
finally leads to the critical PDE for the real functions }${\large u}$%
{\large \ and }${\large v}$%
\begin{eqnarray*}
\mathbf{0} &=&{\large \nabla }u\times {\large \nabla }v=\left( \frac{%
\partial u}{\partial x}\widehat{\mathbf{x}}+\frac{\partial u}{\partial y}%
\widehat{\mathbf{y}}\right) \times \left( \frac{\partial v}{\partial x}%
\widehat{\mathbf{x}}+\frac{\partial v}{\partial y}\widehat{\mathbf{y}}%
\right) = \\
&=&\frac{\partial u}{\partial x}\frac{\partial v}{\partial y}\left( \widehat{%
\mathbf{x}}\times \widehat{\mathbf{y}}\right) +\frac{\partial u}{\partial y}%
\frac{\partial v}{\partial x}\left( \widehat{\mathbf{y}}\times \widehat{%
\mathbf{x}}\right) =\left( \frac{\partial u}{\partial x}\frac{\partial v}{%
\partial y}-\frac{\partial u}{\partial y}\frac{\partial v}{\partial x}%
\right) \left( \widehat{\mathbf{x}}\times \widehat{\mathbf{y}}\right)
\end{eqnarray*}%
{\large or the equation}%
\begin{equation}
\left\{ \frac{\partial u}{\partial x}\frac{\partial v}{\partial y}-\frac{%
\partial u}{\partial y}\frac{\partial v}{\partial x}\right\} \left(
x,y\right) =0\text{ \ \ }\forall (x,y)\in \Omega _{\alpha }\cap \Omega
_{\beta }  \tag{3.10}  \label{3.10}
\end{equation}%
{\large At this point, it's clear that the differential equation (3.10) is
the unique necessary condition for an arbitrary complex flux function }$\Psi 
${\large \ so the general free fields transformations (3.1),(3.2) are a
mathematical coherent choice for the equilibrium problem in object;
furthermore, it's remarkable that this condition doesn't imply the
analiticity of the function }$\Psi $; {\large indeed, if the analiticity is
requested in }$\Omega _{\alpha }\cap \Omega _{\beta }$ {\large for }$\Psi $,%
{\large \ the functions }${\large u}${\large \ and }${\large v}$ {\large %
must be }$%
\mathbb{C}
$ {\large - differentiable in the domain }$\Lambda =\overline{\left( \Omega
_{\alpha }\cap \Omega _{\beta }\right) }\setminus \partial \left( \Omega
_{\alpha }\cap \Omega _{\beta }\right) $, {\large or} {\large the
Cauchy-Riemann conditions must be valid in }$\Lambda $; {\large these
deliver for }${\large u}$%
\begin{equation}
\left\{ \left( \frac{\partial u}{\partial x}\right) ^{2}+\left( \frac{%
\partial u}{\partial y}\right) ^{2}\right\} \left( x,y\right) =0\text{ \ \ }%
\forall (x,y)\in \Lambda  \tag{3.11}  \label{3.11}
\end{equation}%
{\large while for }${\large v}$ {\large we have similarly}%
\begin{equation}
\left\{ \left( \frac{\partial v}{\partial x}\right) ^{2}+\left( \frac{%
\partial v}{\partial y}\right) ^{2}\right\} \left( x,y\right) =0\text{ \ \ }%
\forall (x,y)\in \Lambda  \tag{3.12}  \label{3.12}
\end{equation}%
{\large and it's clear that the differential equations (3.11) and (3.12) are
specific cases which are included in the differential equation (3.10).}

{\large Usefully}, {\large we note that if the complex flux function }$\Psi $%
{\large \ is in the Gauss form}%
\begin{equation}
\Psi (x,y)=\rho (x,y)e^{i\phi (x,y)}\text{ \ \ \ }\forall (x,y)\in \Omega
_{\alpha }\cap \Omega _{\beta }  \tag{3.13}  \label{3.13}
\end{equation}%
{\large the critical condition (3.9) must return}%
\begin{equation}
{\large \nabla }\rho \times {\large \nabla }\phi =\mathbf{0}\text{ \ \ }%
\forall (x,y)\in \Omega _{\alpha }\cap \Omega _{\beta }  \tag{3.14}
\label{3.14}
\end{equation}%
{\large Let us observe that if we impose the analiticity of the flux
function in the equilibria domain }$\left( \Omega _{\alpha }\cap \Omega
_{\beta }\right) ^{\circ }${\large , the Helmholtz problem (2.27)}%
\begin{equation}
\text{\ }{\Large \nabla }^{2}\Psi +{\Large k}\left( \Psi \right) =0\text{ \
\ \ }\forall (x,y)\in \Omega _{\alpha }\cap \Omega _{\beta }  \notag
\end{equation}%
{\large deliver the functional equations system}%
\begin{eqnarray}
{\Large \nabla }^{2}\Psi &=&0\text{ \ \ \ }  \TCItag{3.14.1}  \label{3.14.1}
\\
{\Large k}\left( \Psi \right) &=&0\text{ \ \ }\forall (x,y)\in \Omega
_{\alpha }\cap \Omega _{\beta }  \TCItag{3.14.2}  \label{3.14.2}
\end{eqnarray}%
{\large these equations destroy obviously the related NLSE-problem but they
represents also a not-banal question from the closely functional point of
view, because the equation (3.14.2) must be satisfied only in a 'sub-domain' 
}$\Omega _{\alpha }\cap \Omega _{\beta }$ {\large of the entire domain }$%
Dom\left( \Psi \right) $; {\large indeed} {\large these domains} {\large are
subject to the condition}%
\begin{equation*}
\left( \Omega _{\alpha }\cap \Omega _{\beta }\right) \subseteq Dom\left(
\Psi \right)
\end{equation*}%
{\large leaving free in this way the choice of the not-bound portion of }$%
\Psi ${\large \ inside }$Dom\left( \Psi \right) \setminus \left( \Omega
_{\alpha }\cap \Omega _{\beta }\right) $. {\large This means that a certain
family of }$\Psi $ {\large functions\ which satisfy the equation (3.14.2)
can be identified as}%
\begin{equation*}
\widetilde{\Psi }=h\left( \Omega _{\alpha }\cap \Omega _{\beta }\right)
=\Psi \text{ \ \ }\forall (x,y)\in \left( \Omega _{\alpha }\cap \Omega
_{\beta }\right) \cap Dom\left( \Psi \right)
\end{equation*}%
{\large \ and it means also that this family must depend only on the
possible forms of the flux functions inside the complementary domain }$%
Dom\left( \Psi \right) \setminus \left( \Omega _{\alpha }\cap \Omega _{\beta
}\right) $.{\large \ Now, because the (3.14.1), we can note that if }$\left(
\Omega _{\alpha }\cap \Omega _{\beta }\right) \neq ${\large \ }$Dom\left(
\Psi \right) $, {\large it can be possible to define a family of harmonic
functions }$\widetilde{\Psi }$ {\large inside }$\Omega _{\alpha }\cap \Omega
_{\beta }$ {\large which satisfy the functional equation (3.14.2); this
functions family is related obviously to the solutions set of the equation}%
\begin{equation}
{\large g}\left( \left\vert \Psi \right\vert ^{2}\right) {\large +i\sigma }%
\left( \left\vert \Psi \right\vert ^{2}\right) =0\text{ \ \ }\forall
(x,y)\in \Omega _{\alpha }\cap \Omega _{\beta }  \tag{3.14.3}  \label{3.14.3}
\end{equation}%
{\large At this point,} {\large it's remarkable that this equation
necessarily delivers the above not-trivial solution for }$\Psi $,{\large \
because there is the close condition}%
\begin{equation}
\left( \Omega _{\alpha }\cap \Omega _{\beta }\right) \subset Dom\left( \Psi
\right)  \tag{3.14.4}  \label{3.14.4}
\end{equation}
{\large which} {\large can be justified by observing that the equilibria
domain does not concern in general the entire framework of the equatorial
plane in the accretion disc unlike }$\Psi ${\large .}

{\large Let us note that} {\large in our specific case the solutions set of
the (3.14.3) is not trivial because the general equation scenario (3.14.3)
assume the particular form}%
\begin{equation*}
{\large g}\left( \left\vert \Psi \right\vert ^{2},u,v\right) {\large %
+i\sigma }\left( \left\vert \Psi \right\vert ^{2},u,v\right) =0\text{ \ \ }%
\forall (x,y)\in \Omega _{\alpha }\cap \Omega _{\beta }
\end{equation*}%
{\large related to the expressions (2.43a)-(2.44b). It's clear that this
case is generally independent of the condition (3.14.4).}

{\large It's important too, in conclusion, to observe that the 'knotty
structures' can be present only inside a sub-domain }$\Omega _{\alpha }\cap
\Omega _{\beta }{\large \ }${\large for which the condition (3.14.4) is
valid,} {\large giving sense in this way to the plasma equilibria in the
related space regions; hence, it's necessary that the contour conditions for
the plasma equilibrium-dependent differential Helmholtz problem}%
\begin{equation*}
{\Large \nabla }^{2}\Psi +\left( {\large g}\left( \left\vert \Psi
\right\vert ^{2}\right) {\large +i\sigma }\left( \left\vert \Psi \right\vert
^{2}\right) \right) \Psi =0
\end{equation*}%
{\large must be specified on the contours of an appropriate domain }$\Omega $%
{\large , taking for this }$\Omega \subseteq \Omega _{\alpha }\cap \Omega
_{\beta }$.

\begin{notation}
{\large The critical condition (3.9) is identically satisfied for the }%
\textit{Throumoulopoulos et al.}{\large \ transformations \cite{9}, while
for the }\textit{Lapenta}{\large \ transformations \cite{8} we have that the
above condition remains valid in its general form, leading to a free choice
for the real flux functions }${\large u}${\large \ and }${\large v}$, 
{\large which in this case are related only to the bond equation (3.10).
This is because in the second choice is present a term equal to }$\overline{%
\Psi }$, {\large while this one there is not in the first choice. Such
situation means that both choices represent coherent transformation models
but the }\textit{Throumoulopoulos et al. {\large choice} {\large is
mathematically more strong than }Lapenta }{\large choice; from this
consideration clearly doesn't derive that the choice (2.45a),(2.45b) is less
valid than the choice (2.46a),(2.46b). It's important instead that this
observation reconciles in this way the results of the above authors.}
\end{notation}

\begin{notation}
{\large One of the solutions for the equation (3.6) is}%
\begin{equation*}
J\left( \Psi ,\left\vert \Psi \right\vert ^{2}\right) =0\text{ \ \ \ }%
\forall (x,y)\in \Omega _{\alpha }\cap \Omega _{\beta }
\end{equation*}%
{\large Let us note that this functional equation has not a trivial solution
if we consider an appropriate family of flux functions }$\widehat{\Psi }$ 
{\large as we did above in a similar problem; in this way, we could obtain
another valid critical condition on the }$\Psi $ {\large function for the
general fields transformations. An idea is to study the character of the
functional power series}%
\begin{equation*}
J_{\infty }\left( \Psi ,\left\vert \Psi \right\vert ^{2}\right) =\underset{%
i,j}{\text{ }\overset{\infty }{\sum }}c_{ij}\Psi ^{j+1}i\overline{\Psi }%
^{i-1}
\end{equation*}%
{\large for an appropriate choice of the constants set }$\left\{
c_{ij}\right\} _{i,j\in \text{ }%
\mathbb{N}
},$ $c_{ij}\in 
\mathbb{C}
$, {\large taking into account that it must be}%
\begin{equation*}
J_{\infty }\left( \Psi ,\left\vert \Psi \right\vert ^{2}\right) =0\text{ \ \ 
}\forall (x,y)\in \Omega _{\alpha }\cap \Omega _{\beta }
\end{equation*}
\end{notation}

\subsection{On a reality condition for the free fields: physical aspects}

\bigskip {\large \ In previous section we talked about a plausible physical
validity condition for the general choice (3.1),(3.2), saying that the
'reality' of the terms proportional to the magnetic force density and the
pure hydrodynamical pressure force density inside it can be a fundamental
condition from the physical point of view and it's remarkable that such
condition doesn't affect the complexity of the flux function }$\Psi $; 
{\large indeed these terms are the two 'real' vectorial physical variables,
by the free fields transformations,}%
\begin{equation}
{\large \nabla B}_{z}^{2}=2\underset{i,j}{\sum }a_{ij}\overline{\Psi }%
^{i}\Psi ^{j+1}{\Large \nabla }\Psi   \tag{3.15}  \label{3.15}
\end{equation}%
\begin{equation}
{\large \nabla p}=\underset{i,j}{\sum }b_{ij}\overline{\Psi }^{i}\Psi ^{j+1}%
{\Large \nabla }\Psi   \tag{3.16}  \label{3.16}
\end{equation}%
{\large which must be therefore}%
\begin{equation}
{\large \nabla B}_{z}^{2},{\large \nabla p\notin 
\mathbb{R}
}_{%
\mathbb{C}
}^{3},\text{ \ where \ }{\large 
\mathbb{R}
}_{%
\mathbb{C}
}^{3}=\left\{ \mathbf{x}_{%
\mathbb{C}
}\mathbf{:x}_{%
\mathbb{C}
}=\mathbf{x}+i\mathbf{y},\forall \mathbf{x,y\in }\text{ }{\large 
\mathbb{R}
}^{3}\right\}   \tag{3.17}  \label{3.17}
\end{equation}%
{\large Now, for to determine this reality condition for the above force
densities, we impose}%
\begin{equation}
\func{Im}\left( {\large \nabla B}_{z}^{2}\right) =\func{Im}\left( {\large %
\nabla p}\right) =\mathbf{0}  \tag{3.18}  \label{3.18}
\end{equation}%
{\large and taking the polynomial functional (2.34)}%
\begin{equation}
\underset{i,j}{\sum }\gamma _{ij}\overline{\Psi }^{i}\Psi ^{j}={\large g}%
\left( \left\vert \Psi \right\vert ^{2}\right) {\large +i\sigma }\left(
\left\vert \Psi \right\vert ^{2}\right) ,\text{\ \ }\gamma _{ij}\in 
\mathbb{C}
,\text{ }\forall (x,y)\in \Omega _{\alpha }\cap \Omega \text{ }  \notag
\end{equation}%
{\large for (3.15) and (3.16) we obtain}%
\begin{equation}
{\large \nabla B}_{z}^{2}=2\left[ {\large g}\left( \left\vert \Psi
\right\vert ^{2}\right) {\large +i\sigma }\left( \left\vert \Psi \right\vert
^{2}\right) \right] \Psi {\Large \nabla }\Psi   \tag{3.19}  \label{3.19}
\end{equation}%
\begin{equation}
{\large \nabla p}=\left[ {\large g}\left( \left\vert \Psi \right\vert
^{2}\right) {\large +i\sigma }\left( \left\vert \Psi \right\vert ^{2}\right) %
\right] \Psi {\Large \nabla }\Psi   \tag{3.20}  \label{3.20}
\end{equation}%
{\large which are different only for a factor }$2$. {\large At this point,
for the validity of both (3.18) relations, it's sufficient obviously to
analyze the condition}%
\begin{equation}
\func{Im}\left\{ \left[ {\large g}\left( \left\vert \Psi \right\vert
^{2}\right) {\large +i\sigma }\left( \left\vert \Psi \right\vert ^{2}\right) %
\right] \Psi {\Large \nabla }\Psi \right\} =\mathbf{0}  \tag{3.21}
\label{3.21}
\end{equation}%
{\large For this one, taking in account that}%
\begin{equation*}
\Psi (x,y)=u(x,y)+iv(x,y)\text{ \ \ \ }\forall (x,y)\in \Omega _{\alpha
}\cap \Omega _{\beta }
\end{equation*}%
{\large let us calculate in clear complex form the expression }%
\begin{equation*}
\left[ {\large g}\left( \left\vert \Psi \right\vert ^{2}\right) {\large %
+i\sigma }\left( \left\vert \Psi \right\vert ^{2}\right) \right] \Psi 
{\Large \nabla }\Psi 
\end{equation*}%
{\large We have therefore\ }%
\begin{equation*}
\left( {\large g+i\sigma }\right) \Psi {\Large \nabla }\Psi =\left( {\large %
g+i\sigma }\right) \left( {\large u+iv}\right) {\Large \nabla }\left(
u+iv\right) =
\end{equation*}%
\begin{eqnarray*}
&=&\left[ \left( {\large gu-\sigma v}\right) {\large +i}\left( {\large %
gv+\sigma u}\right) \right] \left( {\large \nabla }u+i{\large \nabla }%
v\right) = \\
&=&\left[ \left( {\large gu-\sigma v}\right) {\large \nabla }u-\left( 
{\large gv+\sigma u}\right) {\large \nabla }v\right] +i\left[ \left( {\large %
gu-\sigma v}\right) {\large \nabla }v+\left( {\large gv+\sigma u}\right) 
{\large \nabla }u\right] 
\end{eqnarray*}%
{\large which delivers for the condition (3.21)}%
\begin{equation*}
\left( {\large gu-\sigma v}\right) {\large \nabla }v+\left( {\large %
gv+\sigma u}\right) {\large \nabla }u=\mathbf{0}
\end{equation*}%
{\large or}%
\begin{equation}
{\large \nabla }u=\frac{\left( {\large \sigma v-gu}\right) }{\left( {\large %
gv+\sigma u}\right) }{\large \nabla }v\text{ \ \ }\forall (x,y)\in \Omega
_{\alpha }\cap \Omega _{\beta }  \tag{3.22}  \label{3.22}
\end{equation}%
{\large where the form of the functions }${\large \sigma }$ {\large and }$%
{\large g}$\ {\large is derived by the relations (2.43a)-(2.44b); it's clear
that (3.22) represents a necessary and sufficient condition for (3.21) and
furthermore it leads to the 'new' critical condition}%
\begin{equation}
{\large \nabla }u\times {\large \nabla }v=\mathbf{0}\text{ \ \ }\forall
(x,y)\in \Omega _{\alpha }\cap \Omega _{\beta }  \tag{3.23}  \label{3.23}
\end{equation}%
{\large It's important to observe that this one is the same of the critical
condition from the mathematical point of view, which can be found in (3.9); 
\textit{this means that the critical condition (3.23) ia a 'common necessary
condition' for both the mathematical and physical aspects of the coherence
for the transformation functionals (3.15),(3.16)}.}

{\large \ Now, if we note that the generic solution choosed in a plane
framework }$(x,y)${\large \ for the magnetic induction field is (by the
(2.4)) }$\forall (x,y)\in \Omega _{\alpha }\cap \Omega _{\beta }${\large \ }%
\begin{equation*}
\mathbf{B}=\widehat{\mathbf{z}}\times {\Large \nabla }\Psi +B_{z}\widehat{%
\mathbf{z}}
\end{equation*}%
{\large is reasonable to think that it must be for a complex }$\Psi $, 
{\large as in the conditions (3.18), } 
\begin{equation}
\func{Im}\left( \mathbf{B}\right) =\mathbf{0}\text{ \ \ }\forall (x,y)\in
\Omega _{\alpha }\cap \Omega _{\beta }  \tag{3.24}  \label{3.24}
\end{equation}%
{\large which delivers necessarily}%
\begin{eqnarray*}
\mathbf{0} &\mathbf{=}&\func{Im}\left( \widehat{\mathbf{z}}\times {\Large %
\nabla }\Psi +B_{z}\widehat{\mathbf{z}}\right) =\func{Im}\left( \widehat{%
\mathbf{z}}\times {\Large \nabla }\Psi \right) =\func{Im}\left( \widehat{%
\mathbf{z}}\times \left( {\large \nabla }u+i{\large \nabla }v\right) \right)
= \\
&=&\func{Im}\left( \widehat{\mathbf{z}}\times {\large \nabla }u+i\widehat{%
\mathbf{z}}\times {\large \nabla }v\right) =\widehat{\mathbf{z}}\times 
{\large \nabla }v
\end{eqnarray*}%
{\large or}%
\begin{equation}
\widehat{\mathbf{z}}\times {\large \nabla }v=\mathbf{0}  \tag{3.25}
\label{3.25}
\end{equation}%
{\large this condition leads to below vectorial differential equation by the
expansion of the operator }${\large \nabla }$%
\begin{equation*}
\widehat{\mathbf{z}}\times \left( \frac{\partial v}{\partial x}\widehat{%
\mathbf{x}}+\frac{\partial v}{\partial y}\widehat{\mathbf{y}}\right) =\frac{%
\partial v}{\partial x}\widehat{\mathbf{y}}-\frac{\partial v}{\partial y}%
\widehat{\mathbf{x}}=\mathbf{0}
\end{equation*}%
{\large which means}%
\begin{equation}
\frac{\partial v}{\partial x}=\frac{\partial v}{\partial y}=0\text{ \ \ or \
\ }{\large \nabla }v=\mathbf{0}\text{ \ \ }\forall (x,y)\in \Omega _{\alpha
}\cap \Omega _{\beta }  \tag{3.26}  \label{3.26}
\end{equation}%
{\large At this point it's indispensable to observe that if the imaginary
component of the flux function }$\Psi $ {\large has a zero gradient in }$%
\Omega _{\alpha }\cap \Omega _{\beta }$, {\large the critical common
condition (3.23) is identically verified; so, the only position which can
respect a reality condition for all fields in the equilibrium problem, as
viewed, is therefore a unique flux function condition for which it must be}%
\begin{equation}
\func{Im}\left( \mathbf{\Psi }(x,y)\right) =v(x,y)=\text{cost \ \ }\forall
(x,y)\in \Omega _{\alpha }\cap \Omega _{\beta }  \tag{3.27}  \label{3.27}
\end{equation}%
{\large which delivers finally}%
\begin{equation}
\Psi (x,y)=u(x,y)+iA,\text{ \ }A\in 
\mathbb{R}
,\text{\ \ }\forall (x,y)\in \Omega _{\alpha }\cap \Omega _{\beta } 
\tag{3.28}  \label{3.28}
\end{equation}%
{\large \ \ }

\subsection{On a global critical condition for the GS equilibrium problem:\
critical equations system}

{\large Here we briefly explore a further validity condition for the
acceptability of the general choice (3.1),(3.2). This condition regards the
equation (2.8), which leads to a final 'validity equation' if it's related
to the common critical condition (3.9) or to the flux function condition
(3.27), taking into account the relations (2.11) and (2.14).}

{\large If we start from the equation (2.8), we obtain by applying the
operator }${\large \nabla \times }${\large \ to both sides of this one}%
\begin{eqnarray*}
\mathbf{0}_{%
\mathbb{C}
} &\mathbf{=}&{\large \nabla \times }\left( {\Large \nabla }{\large p\ }+%
{\Large \nabla }^{2}\Psi {\Large \nabla }\Psi +B_{z}{\Large \nabla }%
B_{z}\right) ={\Large \nabla }{\large \times }{\Large \nabla }{\large p\ }+%
{\Large \nabla }{\large \times }{\Large \nabla }^{2}\Psi {\Large \nabla }%
\Psi +{\Large \nabla }{\large \times }B_{z}{\Large \nabla }B_{z}= \\
&=&{\Large \nabla }^{2}\Psi {\Large \nabla }{\large \times }{\Large \nabla }%
\Psi -{\Large \nabla }\Psi \times {\Large \nabla \nabla }^{2}\Psi +B_{z}%
{\Large \nabla }{\large \times }{\Large \nabla }B_{z}-{\Large \nabla }%
B_{z}\times {\Large \nabla }B_{z}= \\
&=&-{\Large \nabla }\Psi \times {\Large \nabla \nabla }^{2}\Psi ={\Large %
\nabla \nabla }^{2}\Psi \times {\Large \nabla }\Psi
\end{eqnarray*}%
{\large or}%
\begin{equation}
{\Large \nabla \nabla }^{2}\Psi \times {\Large \nabla }\Psi =\mathbf{0}_{%
\mathbb{C}
}  \tag{3.29}  \label{3.29}
\end{equation}%
{\large This further condition is obviously a necessary condition for the
Grad-Shafranov equilibrium problem and it can be translate into below
differential equations set: if }$\Psi $ {\large is the complex function }$%
\Psi (x,y)=u(x,y)+iv(x,y)$ \ \ \ $\forall (x,y)\in \Omega _{\alpha }\cap
\Omega _{\beta }$, {\large we have }%
\begin{eqnarray*}
\mathbf{0}_{%
\mathbb{C}
} &=&{\large \nabla \nabla }^{2}\Psi {\large \times \nabla \Psi =\nabla
\nabla }^{2}\left( u+iv\right) {\large \times \nabla }\left( u+iv\right) = \\
&=&\left( {\large \nabla \nabla }^{2}u+i{\large \nabla \nabla }^{2}v\right)
\times \left( {\large \nabla }u+i{\large \nabla }v\right) = \\
&=&{\large \nabla \nabla }^{2}u\times {\large \nabla }u-{\large \nabla
\nabla }^{2}v\times {\large \nabla }v+i\left( {\large \nabla \nabla }%
^{2}u\times {\large \nabla }v+{\large \nabla \nabla }^{2}v\times {\large %
\nabla }u\right)
\end{eqnarray*}%
{\large which delivers the 'critical equations system'}%
\begin{eqnarray}
{\large \nabla \nabla }^{2}u\times {\large \nabla }u-{\large \nabla \nabla }%
^{2}v\times {\large \nabla }v &=&\mathbf{0}  \TCItag{3.30a}  \label{3.30a} \\
{\large \nabla \nabla }^{2}u\times {\large \nabla }v+{\large \nabla \nabla }%
^{2}v\times {\large \nabla }u &=&\mathbf{0}  \TCItag{3.30b}  \label{3.30b}
\end{eqnarray}%
{\large Now, if we consider the flux function condition (3.27), we have} 
{\large that the second equation is identically verified, while for the
first equation we obtain}%
\begin{equation}
{\large \nabla \nabla }^{2}u\times {\large \nabla }u=\mathbf{0}\text{ \ \ }%
\forall (x,y)\in \Omega _{\alpha }\cap \Omega _{\beta }  \tag{3.31}
\label{3.31}
\end{equation}%
{\large or}%
\begin{eqnarray*}
\mathbf{0} &=&{\large \nabla \nabla }^{2}u\times {\large \nabla }u=\left( 
\frac{\partial }{\partial x}\widehat{\mathbf{x}}+\frac{\partial }{\partial y}%
\widehat{\mathbf{y}}\right) {\large \nabla }^{2}u\times \left( \frac{%
\partial u}{\partial x}\widehat{\mathbf{x}}+\frac{\partial u}{\partial y}%
\widehat{\mathbf{y}}\right) = \\
&=&\frac{\partial }{\partial x}{\large \nabla }^{2}u\frac{\partial u}{%
\partial y}\left( \widehat{\mathbf{x}}\times \widehat{\mathbf{y}}\right) +%
\frac{\partial }{\partial y}{\large \nabla }^{2}u\frac{\partial u}{\partial x%
}\left( \widehat{\mathbf{y}}\times \widehat{\mathbf{x}}\right) = \\
&=&\left( \frac{\partial }{\partial x}{\large \nabla }^{2}u\frac{\partial u}{%
\partial y}-\frac{\partial }{\partial y}{\large \nabla }^{2}u\frac{\partial u%
}{\partial x}\right) \left( \widehat{\mathbf{x}}\times \widehat{\mathbf{y}}%
\right)
\end{eqnarray*}%
{\large which delivers the unique differential equation for all critical
conditions}%
\begin{equation}
\left( \frac{\partial }{\partial x}{\large \nabla }^{2}u\frac{\partial u}{%
\partial y}-\frac{\partial }{\partial y}{\large \nabla }^{2}u\frac{\partial u%
}{\partial x}\right) =0\text{ \ \ }\forall (x,y)\in \Omega _{\alpha }\cap
\Omega _{\beta }  \tag{3.32}  \label{3.32}
\end{equation}%
{\large If we consider instead the common critical condition (3.9) only,}%
\begin{equation*}
{\large \nabla }u\times {\large \nabla }v=\mathbf{0}\text{ \ \ }\forall
(x,y)\in \Omega _{\alpha }\cap \Omega _{\beta }
\end{equation*}%
{\large we obtain by the system (3.30a),(3.30b)}%
\begin{equation}
{\large \nabla \nabla }^{2}u\times {\large \nabla }u=\mathbf{0}\text{ \ \ }%
\forall (x,y)\in \Omega _{\alpha }\cap \Omega _{\beta }  \tag{3.33}
\label{3.33}
\end{equation}%
{\large which delivers, as in (3.32), the differential equation}%
\begin{equation}
\left( \frac{\partial }{\partial x}{\large \nabla }^{2}v\frac{\partial v}{%
\partial y}-\frac{\partial }{\partial y}{\large \nabla }^{2}v\frac{\partial v%
}{\partial x}\right) =0\text{ \ \ }\forall (x,y)\in \Omega _{\alpha }\cap
\Omega _{\beta }  \tag{3.34}  \label{3.34}
\end{equation}%
{\large taking into account that the condition (3.9) means}%
\begin{equation*}
{\large \nabla }u=f\left( u,v\right) {\large \nabla }v
\end{equation*}%
{\large where }$f$ {\large is a real functional on }$u$ {\large and}\ $v$,\ $%
\forall (x,y)\in \Omega _{\alpha }\cap \Omega _{\beta }$.

{\large At this point, it's clear that the conditioning equations (3.32) and
(3.34) represent two \textit{global critical conditions} for the
magnetohydrodynamical equilibrium problem in the pseudo-general NLSE
framework, for which obviously it's true that}%
\begin{equation}
\left( \frac{\partial }{\partial x}{\large \nabla }^{2}u\frac{\partial u}{%
\partial y}-\frac{\partial }{\partial y}{\large \nabla }^{2}u\frac{\partial u%
}{\partial x}\right) =0\Longrightarrow  \tag{3.35}  \label{3.35}
\end{equation}%
\begin{equation*}
\Longrightarrow \left( \frac{\partial }{\partial x}{\large \nabla }^{2}v%
\frac{\partial v}{\partial y}-\frac{\partial }{\partial y}{\large \nabla }%
^{2}v\frac{\partial v}{\partial x}\right) =0\text{\ \ \ }\forall (x,y)\in
\Omega _{\alpha }\cap \Omega _{\beta }
\end{equation*}%
{\large for the bond equations system (3.30a),(3.30b).}

\section{Comments}

{\large In conclusion: we have seen in previous sections that for to obtain
a pseudo-general form of the NLSE in the stationary case (see (2.22) in the
plane framework }$(x,y)${\large ) by the GSEs set (see the system
(2.7),(2.8)) for a general plasma equilibrium problem in the equatorial
plane of an accretion disc, or}%
\begin{equation*}
{\Large \nabla }^{2}\Psi +\left( {\large g}\left( \left\vert \Psi
\right\vert ^{2}\right) {\large +i\sigma }\left( \left\vert \Psi \right\vert
^{2}\right) \right) \Psi =0
\end{equation*}%
{\large a set of transformation relations (i.e. the general choice) of the
free fields in the GSE is imposable in a general form and this is}%
\begin{equation*}
{\large B}_{z}{\large \nabla B}_{z}=\frac{1}{2}{\large \nabla B}_{z}^{2}=%
\underset{i,j}{\sum }a_{ij}\overline{\Psi }^{i}\Psi ^{j+1}{\Large \nabla }%
\Psi
\end{equation*}%
\begin{equation*}
{\large \nabla p}=\underset{i,j}{\sum }b_{ij}\overline{\Psi }^{i}\Psi ^{j+1}%
{\Large \nabla }\Psi \text{ \ \ \ \ {\large with \ \ \ }}a_{ij},b_{ij}\in 
\mathbb{C}%
\end{equation*}%
{\large where }$\Psi $ {\large is the flux function in the solution (2.4) of
the general equilibrium problem (2.3) and it's a complex function on the
real variables }$x$ {\large and} $y$%
\begin{equation*}
\Psi (x,y)=u(x,y)+iv(x,y)\text{ \ \ \ }\forall (x,y)\in \Omega _{\alpha
}\cap \Omega _{\beta }
\end{equation*}%
{\large which is a solution for the Helmholtz problem (2.25) and it's
defined in a specific local not-banal domain (in the accretion disc
framework) as }$\Omega _{\alpha }\cap \Omega _{\beta }$; {\large it's clear
that the topology of this domain is a very fundamental characteristic of the
equilibrium problem, because the critical condition equation (3.10) must be
verified for the mathematical coherence of the above fields transformations
inside it, or}%
\begin{equation*}
\left\{ \frac{\partial u}{\partial x}\frac{\partial v}{\partial y}-\frac{%
\partial u}{\partial y}\frac{\partial v}{\partial x}\right\} \left(
x,y\right) =0\text{ \ \ }\forall (x,y)\in \Omega _{\alpha }\cap \Omega
_{\beta }
\end{equation*}%
{\large which, in the vectorial form, is}%
\begin{equation*}
{\large \nabla }u\times {\large \nabla }v=\mathbf{0}\text{ \ \ }\forall
(x,y)\in \Omega _{\alpha }\cap \Omega _{\beta }
\end{equation*}%
{\large and therefore it's remarkable that the analiticity of the flux
function is not necessary for the validity of the general fields choice.
Furthermore, this equation represents a global critical condition for the
equilibrium problem from the solitonic point of view, as showed in (3.35).
We remember that the critical condition (3.10) is equal to the condition on }%
$\Psi $%
\begin{equation*}
{\large \nabla }\Psi \times {\large \nabla }\overline{\Psi }=\mathbf{0}\text{
\ \ }\forall (x,y)\in \Omega _{\alpha }\cap \Omega _{\beta }
\end{equation*}

{\large It's clear at this point that the }\textit{Throumoulopoulos et al.}%
{\large \ and }\textit{Lapenta}{\large \ theoretical positions (see
(2.45a-b) and (2.46a-b)) about a specific transformations set for the free
fields are cannot opposite, because both these choices satisfy the critical
condition (3.8b), which is identically verified for the first choice, while
it is in its general form in the second case because however it remains
verified for an appropriate choice of the }$u$ {\large and}\ $v$ {\large real%
} {\large functions inside}\ $\Omega _{\alpha }\cap \Omega _{\beta }$.

\end{document}